\documentstyle[eqsecnum,aps,epsfig,psfig,12pt]{revtex}
\def\di{\displaystyle}
\def\tab{&\di}
\newcommand{\step}{\vspace{.5em}}

\def\eq#1{(\ref{#1})}
\def\Eq#1{Eq.~(\ref{#1})}
\def\Es#1{Eqs.~(\ref{#1})}

\def\Tr{{\rm Tr}}

\def\ov{\over}
\def\s0#1#2{\mbox{\small{$ \frac{#1}{#2} $}}}
\def\0#1#2{\frac{#1}{#2}}

\def\eq#1{(\ref{#1})}
\def\Eq#1{Eq.~(\ref{#1})}

\def\ad{{\rm ad}}

\date{\today}

\makeatletter

\renewenvironment{thebibliography}[1]
         {\section*{References}\frenchspacing\small
          \begin{list}{[\arabic{enumi}]}
         {\usecounter{enumi}\parsep=2pt\topsep 0pt
         \settowidth{\labelwidth}{[#1]}
         \leftmargin=\labelwidth\advance\leftmargin\labelsep
         \rightmargin=0pt\itemsep=0pt\sloppy}}{\end{list}}

\makeatother

\begin{document}
\begin{center}

\thispagestyle{empty}
\begin{flushright}
{CERN-TH-2002-45\\ FAU-TP3-02-05}
\end{flushright}
\vspace{2cm}

\mbox{
\large 
\bf 
Completeness and consistency of renormalisation group flows
}
\\[6ex]

{Daniel F. Litim
\footnote{Daniel.Litim@cern.ch}
and 
Jan M.~Pawlowski 
\footnote{jmp@theorie3.physik.uni-erlangen.de}}
\\[4ex]
{${}^*${\it 
Theory Division, 
CERN\\
CH-1211 Geneva 23.
}\\[2ex]${}^\dagger${\it 
Institut f\"ur Theoretische Physik III\\ 
Universit\"at Erlangen, 
D-91054 Erlangen.
}}
\\[10ex]
 
{\small \bf Abstract}\\[2ex]
\begin{minipage}{14cm}{\small 
    We study different renormalisation group flows for scale dependent
    effective actions, including exact and proper-time renormalisation
    group flows. These flows have a simple one loop structure. They
    differ in their dependence on the full field-dependent propagator,
    which is linear for exact flows. We investigate the inherent
    approximations of flows with a non-linear dependence on the
    propagator. We check explicitly that standard perturbation theory
    is not reproduced. We explain the origin of the discrepancy by
    providing links to exact flows both in closed expressions and in
    given approximations. We show that proper-time flows are
    approximations to Callan-Symanzik flows. Within a background field
    formalism, we provide a generalised proper-time flow, which is
    exact.  Implications of these findings are discussed.}
\end{minipage}
\end{center}

\newpage
\pagestyle{plain}
\setcounter{page}{1}

\section{Introduction}

Renormalisation group (RG) methods are an essential ingredient in the
study of non-perturbative problems in continuum and lattice
formulations of quantum field theory. Over the last decade increasing
interest has been devoted to particular formulations of RG flows,
which have one main property in common: they all can be written as a
simple one loop equation in the full field-dependent propagator.
Their one loop structure is very useful because it allows to encompass
technical complications due to overlapping loop integrations known
from standard perturbation theory and Schwinger-Dyson equations.
Another important strength of these RG flows
is based on their flexibility, when it comes to truncations of the
full problem under investigation. This makes all the different sets of
RG equations interesting for situations where one has to resort to
approximations because the full problem is to hard to attack. For
non-perturbative effects at strong coupling or large correlation
lengths, such an approach is essentially unavoidable.\step

Despite their close similarity, the various RG flows with a one loop
structure differ qualitatively in important aspects. The RG flows
depend on the precise implementation of a regularisation, typically
given by momentum or operator cutoffs. Furthermore, some RG flows are
known to be {\it exact}, as they can be derived from first principles,
mainly done within a path integral representation. Prominent examples
for such one loop exact flows\footnote{From now on, we refer to
  renormalisation group flows with a one loop structure as ``one loop
  flows''.  Exact flows with a one loop structure are referred to as
  ``one loop exact flows''. This should not to be confused with a one
  loop approximation (i.e.~one loop exact flows are {\it not} one loop
  approximations of some exact flow).} are Exact RG (ERG) flows
\cite{Polchinski:1983gv,CW,Litim:1998nf,Reviews}.  These flows, which
we use as reference points in the present paper, are closely related
to other well-known exact flows like Wilsonian flows \cite{Wilson},
Wegner-Houghton flows \cite{Wegner} and Callan-Symanzik flows
\cite{Callan:1970yg}.  The strength of exact RG flows is that
systematic approximations of the integrated flow correspond to
systematic approximations to the full quantum theory.  This allows to
devise optimisation conditions
\cite{Litim:2000ci,Litim:2001up,Litim:2001dt}, which resolve the
problem of the spurious regulator dependence
\cite{Litim:2000ci,Litim:2001up,Litim:2001dt,Scheme,Freire:2000sx}.\step

In turn, some one loop RG flows have been derived within the
philosophy of a one loop improvement.  This includes the proper-time
RG flows \cite{Liao:1996fp,Liao:1997nm} and RG flows based on an
operator cutoff \cite{Liao:2000yu}.  The similarity between the
different one loop flows with one loop exact flows has fuelled hopes
that the scenario just described for exact flows may be valid in 
general.  Therefore, it is important to either establish that a given
flow is exact, or, if not, what approximation to an exact flow it
represents.  This is at the root for the predictive power of the
formalism. So far, this question has been studied within the
derivative expansion \cite{Litim:2001hk}.  A first account of a more
general analysis was given in \cite{Litim:2001ky}, where we compared
the perturbative expansions of different one loop flows.\step

In the present work we give a general analysis of the problems
mentioned above. A detailed study of the following one loop and one
loop exact flows is provided: ERG flows, Callan-Symanzik flows and
generalisations thereof, proper-time flows, and one loop flows based
on an operator regularisation.  We show that one loop exact flow
depend linearly on the full field-dependent propagator. A general one
loop flow does not have this structure. As a consequence, we show that
integrated non-exact flows deviate from standard perturbation theory
at the first non-trivial order, {\it i.e.}~two loop. Additionally, we
relate proper-time flows to the Callan-Symanzik flow, and -in given
approximations- to ERG flows. We also discuss the possibility of an
exact map between ERG and proper-time flows. Based on these findings,
we present {\it generalised} proper-time flows, which are exact.
\step

It proves helpful to introduce two properties of RG flows which we
refer to as {\it completeness} and {\it consistency}.  Consider a
general flow defined by an initial effective action given at some
initial scale $\Lambda$, and a flow equation connecting it with the
full quantum effective action at vanishing cut-off scale. Then we
define that
\begin{itemize}
\item a flow is {\it consistent}, if its flow equation connects an
  explicitly known initial effective action with the full quantum
  effective action.
\item a {consistent} flow is {\it complete}, if the initial effective
  action is trivial, namely the classical action.
\end{itemize}
As the initial effective action of a complete flow is trivial, all
quantum fluctuations result from integrating the flow equation.
Well-known examples of complete flows are Callan-Symanzik flows and
ERG flows. In turn, for a consistent flow, in general, parts of the
quantum fluctuations are already contained in the initial effective
action. The latter has to be known explicitly.\footnote{This subtlety
  is discussed in Sect.~\ref{PTRG-Consistency}.}  Important examples
for consistent flow are ERG flows with a non-trivial initial effective
action. For thermal field theories, this concerns scenarios where
initial effective actions stem from perturbative dimensional reduction
\cite{Freire:2000sx}, or thermal RG flows within the ERG framework as
provided in \cite{Litim:1998nf} and \cite{D'Attanasio:1997fy}. In the
latter, only thermal fluctuations are integrated by the flow
while the quantum fluctuations have already been integrated out and
are part of the initial effective action. \step

The outline of the paper is as follows.  In Sect.~\ref{SectionERG}, as
an introduction of the methods, we discuss consistency and
completeness for ERG flows.  We argue, that general one loop exact
flows must depend linearly on the full propagator. This result is
derived in App.~\ref{EOL}.  Then we sketch the derivation of ERG flows
from first principles and explicitly show their completeness within
perturbation theory.  Generalisations to consistent ERG flows, in
particular at finite temperature are briefly discussed.

In Sect.~\ref{SectionPTRG} we study proper-time flows.  We sketch
their derivation as one loop improved RG equations. Then we prove that
these flows are in general incomplete. We provide explicit expressions
for the regulator dependent deviation from complete flows at two loop.
We also give their link to Callan Symanzik flows.  It is argued that a
proper-time flow is not a consistent flow.  These findings are
illustrated within a simple example.

In Sect.~\ref{SectionOther}, we discuss consistency and completeness
for flows derived from a multiplicative regularisation of the one loop
momentum integral.  By explicitly calculating the two loop
contributions of the integrated flow we show, that these flows are
neither complete nor consistent.

In Sect.~\ref{SectionCCRG}, we devise maps between given
approximations of proper-time flows and ERG flows.  In addition, we
show how the proper-time regularisation has to be generalised in order
to turn the flow into a complete and consistent flow. This is based on
generalised proper-time regulators and involves the use of the
background field method.

In Sect.~\ref{SectionDiscussion}, we close with a discussion of the
main results and their implications regarding the predictive power of
the different RG flows.

Some more technical aspects are summarised in the appendices. In
App.~\ref{EOL}, it is shown that a general one loop exact flow for the
effective action can only depend linearly on the full propagator.
This result is used at various places in the main body of the work.
In App.~\ref{SectionCS}, we study Callan-Symanzik flows and
generalisations thereof. Proper-time flows can be seen as
approximations to generalised Callan-Symanzik flows. In
App.~\ref{CS-Example}, we compute explicitly the two loop effective
action from a generalised Callan-Symanzik flow. This result is used in
Sect.~\ref{SectionPTRG} for comparison with a specific proper-time
flow. In App.~\ref{secrecursive}, we derive a recursion relation for
the two loop effective action within the standard proper-time RG. This
result is used in Sect.~\ref{SectionPTRG}.  \step

\section{Exact renormalisation group}\label{SectionERG}
In this section we discuss the concepts of completeness and
consistency at the example of ERG equations.  Prior to this, we
comment on the general structure of one loop exact flows. A general
exact flow is the flow of some operator insertion within the
theory. The expectation values of more than two fields involve multi
loop contributions. Thus, insisting on the one loop nature of the flow,
one is bound to an insertion which is at most quadratic in the
fields.  Otherwise, the corresponding exact flow would also contain
higher loop contributions.  We conclude that an exact flow with a one
loop structure must depend {\it linearly} on the full propagator
\cite{Litim:2001ky}. More details are given in App.~\ref{EOL}.
\step

\subsection{Derivation}

The usual starting point is the generating functional of the theory at
hand, where a cut-off term $\Delta S_k[\phi]$ is added to the
classical action. Here we discuss a theory with a scalar field and a
general interaction, the generalisation to arbitrary field content is
straightforward. We have
\begin{equation}
Z_k[J]=
\int d\phi 
\,\exp\left({-S[\phi]-\Delta S_k[\phi]+\int d^d x J\phi}\right),
\end{equation}
where $d$ counts space-time dimensions.  This leads to the
flow equation
\begin{equation}
\partial_t Z_k[J]=-\langle \partial_t \Delta S_k[\phi]\rangle_J\,. 
\label{GF}
\end{equation}
An insertion $\Delta S_k[\phi]$ at most quadratic in the fields
guarantees the one loop structure of the corresponding flow. Hence, we
choose to $\Delta S_k[\phi]=\s012\int d^d x\, \phi\, R\,\phi$, where
$R$ is an infrared (IR) regulator function depending on an IR scale
$k$. Functions $R(q^2)$ have to satisfy a number of conditions in
order to provide an infra-red regularisation for the effective
propagator, and to ensure that the flow \eq{ERG} interpolates between
an initial (classical) action in the UV and the full quantum effective
action in the IR. The necessary conditions on $R_k$ are summarised as
\begin{eqnarray}
\label{1}
\lim_{q^2/k^2\to 0}R_k(q^2)&>&0\\
\label{2}
\lim_{k^2/q^2\to 0}R_k(q^2)&=&0\\
\label{3}
\lim_{k\to \Lambda}R_k(q^2)&\to&\infty\,.
\end{eqnarray} 
where $\Lambda$ is an ultraviolet (UV) scale. Eq.~\eq{1} guarantees
that $R_k$ provides an IR regulator, because massless modes are
effectively cut-off. The second condition \eq{2} ensures that the
regulator is removed in the IR limit $k\to 0$ and that the theory is 
unchanged for momentum modes with $q^2\gg k^2$. The condition \eq{3}
ensures that the correct initial condition is reached for $\lim_{k\to
\Lambda}\Gamma_k=S_\Lambda$. Here, $\Lambda$ is the initial (UV)
scale.  
\step

The effective action is defined as the Legendre transformation
$\Gamma_k[\phi]=\int d^d x J\phi-\ln Z_k[\phi] -\Delta S_k[\phi]$.
This leads to a simple form of the flow equation for $\Gamma_k$. From 
\eq{GF} we get for the flow of the
effective action
\begin{eqnarray} \label{ERG}
\partial_t\Gamma_k[\phi] =
\frac{1}{2}\Tr \left(\Gamma_k^{(2)}+ R_k\right)^{-1}\partial_t  R_k\,,  
\end{eqnarray}
where
\begin{equation}
\Gamma_k^{(2)}[\phi](p,q)=
\0{\delta^2\Gamma_k[\phi]}{\delta\phi(p)\delta\phi(q)}
\end{equation}
and the trace denotes a sum over all momenta and indices, $t=\ln k$.
The ERG flow is linear in the full propagator, as required for an
exact one loop flow. It is IR finite due to \eq{1} and UV finite due
to \eq{2}.

\subsection{Completeness}\label{ERG_Completeness}

It is well-known that perturbation theory is contained in ERG flows.
The first use of this approach was to simplify proofs of perturbative
renormalisability \cite{Polchinski:1983gv}.  The UV boundary condition
$\Gamma_\Lambda$ is the classical action. All quantum fluctuations are
integrated-out along the flow. Therefore, the ERG flow has to be
complete.  An explicit check of completeness is provided by
successively integrating the given flow equation perturbatively order
by order and comparing the result to standard perturbation theory.
Such a check is useful for flows which lack a derivation from first
principles. There, it also provides some insight in the structure of
the deviations.  Here we perform this check for the ERG up to two
loop. It serves as an introduction to the methods used later.\step

In order to simplify the subsequent expressions, we introduce a
short-hand notation by writing $A_{pqrs\cdots}\equiv
A(p,q,r,s\cdots)$ for momentum variables $p,q,r,s,\cdots$, and
repeated indices correspond to a momentum integration
\begin{equation}\label{notation}
A_{qp}B_{pq'}\equiv (A B)_{qq'}
=\int \0{d^dp}{(2\pi)^d}A(q,p)\, B(p,q')\,.
\end{equation}
As an example we rewrite the ERG equation
\eq{ERG} in this notation,
\begin{eqnarray}\label{dWERG} 
\partial_t \Gamma_k =\s012 \left({1\ov \Gamma_k^{(2)}+R}\right)_{pq}
\partial_t R_{qp}\,. 
\end{eqnarray}
A simple graphical representation for \eq{dWERG} is given by Fig.~1.
\begin{figure}
\begin{center}
\unitlength0.001\hsize
\begin{picture}(250,180)
\psfig{file=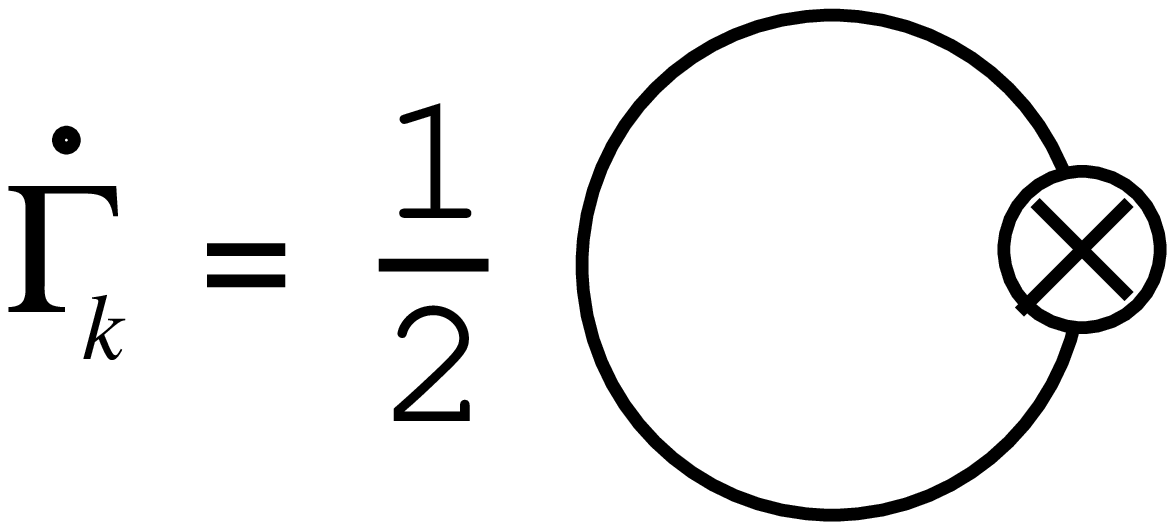,width=.25\hsize}
\end{picture}
\begin{minipage}{.8\hsize}{
{
  \small {\bf Figure 1:} Graphical representation of the ERG equation
  \eq{ERG}.}}
\end{minipage}
\end{center}
\end{figure}

The closed line in Fig.~1 represents the full field-dependent
propagator $(\Gamma^{(2)}[\phi]+R)^{-1}$ and the crossed circle stands
for the insertion $\partial_t R$. According to Fig~1, or \eq{dWERG},
the ERG equation has a simple one loop structure, which should not be
confused with a standard perturbative loop as it contains the full
propagator.  The explicit calculations are most easily carried out
within the graphical representation.  We introduce the graphical
notation as given in Fig.~2.\step

\begin{figure}
\begin{center}
\unitlength0.001\hsize
\begin{picture}(500,220)
\psfig{file=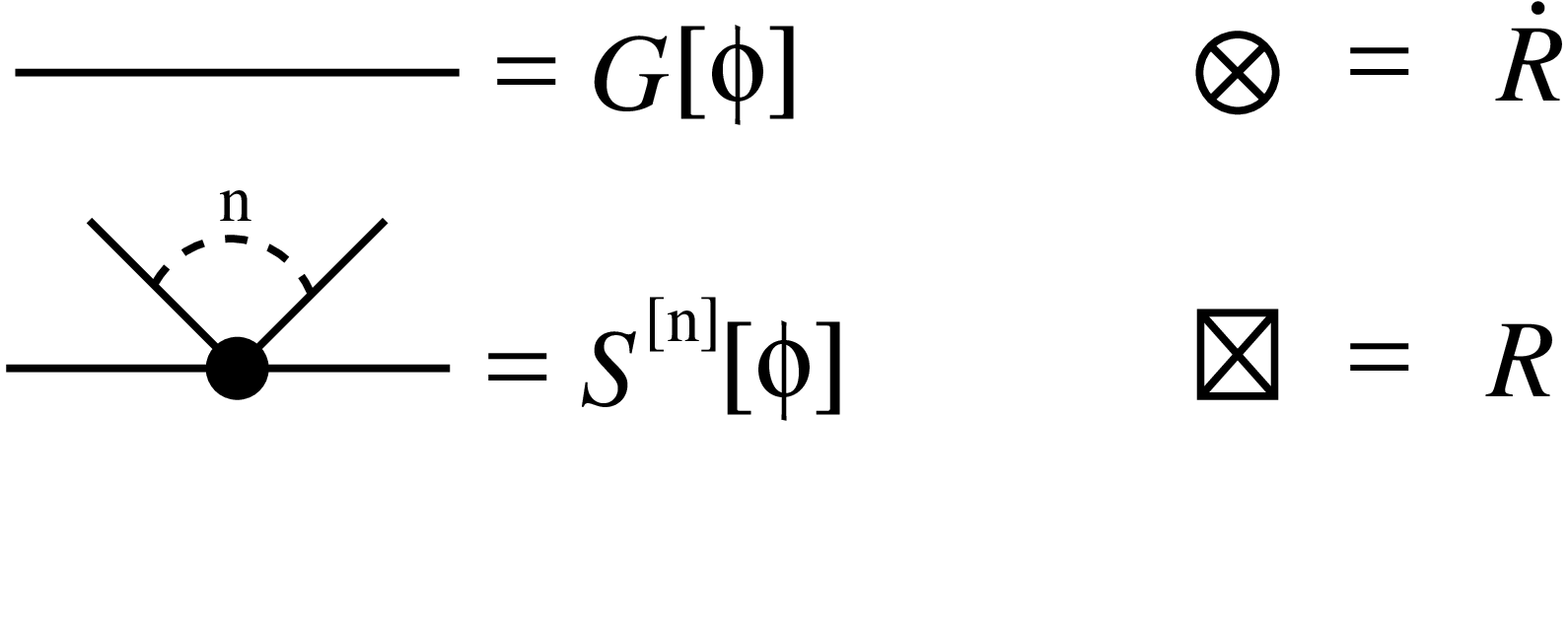,width=.5\hsize}
\end{picture}
\begin{minipage}{.8\hsize}{
    { \small {\bf Figure 2:} Graphical representation of the
      propagator $G[\phi]$, the (classical) $n$-point vertices
      $S^{(n)}[\phi]$, and the insertions $\partial_t R\equiv \dot R$
      and $R$.}  }
\end{minipage}
\end{center}
\end{figure}

The precise expression for the propagator $G[\phi]$ in Fig.~2 depends
on the flow studied. The line in Fig.~2 stands for the field dependent
perturbative propagator $(S^{(2)}[\phi]+R)^{-1}$, in contrast to
Fig.~1.  The vertices are the classical ones, but also with full field
dependence. \step

Now let us write the effective action within a loop expansion 
\begin{eqnarray}
\label{loops}
\Gamma=S+
\sum_{n=1}^\infty\Delta\Gamma_n, 
\end{eqnarray} 
where $S$ is the classical action and $\Delta\Gamma_n$ comprises the
$n$th loop order. At one loop, the integrated flow is
\begin{eqnarray}
\label{ERG1loop} 
\Delta\Gamma_1=\Delta\Gamma_{1,\Lambda}+
\int_\Lambda^k \s0{dk'}{k'} 
\left. \partial_{t'} \Gamma_{k'}\right|_{\rm 1-loop} =
\Delta\Gamma_{1,\Lambda}+\left.
\s012 \left[\ln \left(S^{(2)}+R\right)\right]_{qq}{}\right|_{\Lambda}^k. 
\end{eqnarray}
The expression on the right-hand side of \eq{ERG1loop} approaches the
full one loop effective action for $k\to 0$. The subtraction at
$\Lambda$ provides the necessary UV renormalisation, together with
$\Delta\Gamma_{1,\Lambda}$. As the latter only encodes renormalisation
effects, we drop it from now on. For the two loop calculation we also
need $\Delta\Gamma_1^{(2)}$, which follows from \eq{ERG1loop} as
\begin{eqnarray}
\label{profield1}
\Delta\Gamma_{1,qq'}^{(2)} = 
\s012\, 
\left(  G_{pp'}\ S_{p'pqq'}^{(4)}
      - G_{pp'}\ S_{p'rq}^{(3)}\ G_{rr'}\ S_{r'pq'}^{(3)}
\right)_\Lambda^k\,.
\end{eqnarray}
Again, the indices $q$ and $q'$ stand for the external momenta . Thus,
$\Delta\Gamma_1^{(2)}$ consists of two (subtracted) graphs. Its
graphical representation is given in Fig.~3.  The double lines stand
for subtracted (finite) diagrams. They are introduced in Fig.~4.

\begin{figure}
\begin{center}
\unitlength0.001\hsize
\begin{picture}(500,150)
\psfig{file=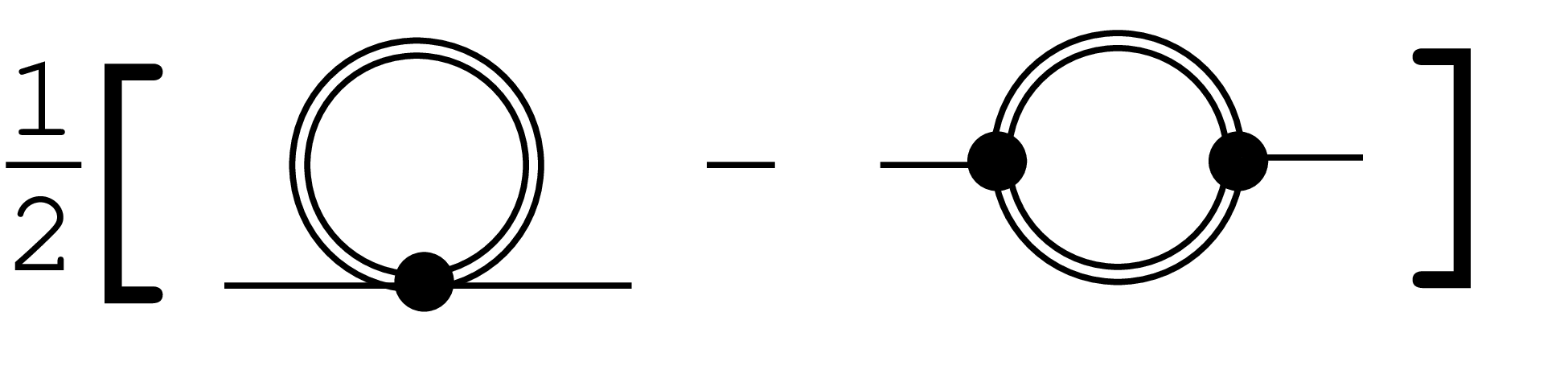,width=.5\hsize}
\end{picture}
\begin{minipage}{.8\hsize}{
{
  \small {\bf Figure 3:} Graphical representation of \eq{profield1}.
  The subtracted diagrams (double lines) are defined in Fig.~4.}}
\end{minipage}
\end{center}
\end{figure}

\noindent 
Clearly the subtraction at $\Lambda$ leads to a renormalisation of the
diagrams. For our purpose these terms are not interesting since they
only provide the details of the renormalisation procedure. Here,
however, we are only interested in the graphical structure of the
perturbation series, including the combinatorial factors.  For this
purpose the structure of the subtractions is irrelevant. In other
words, we want to focus on diagrams, which are evaluated at $k$ even
for subdiagrams. In most results, both graphical and equations, we
will only mention them implicitly.\step

\begin{figure}
\begin{center}
\unitlength0.001\hsize
\begin{picture}(500,350)
\psfig{file=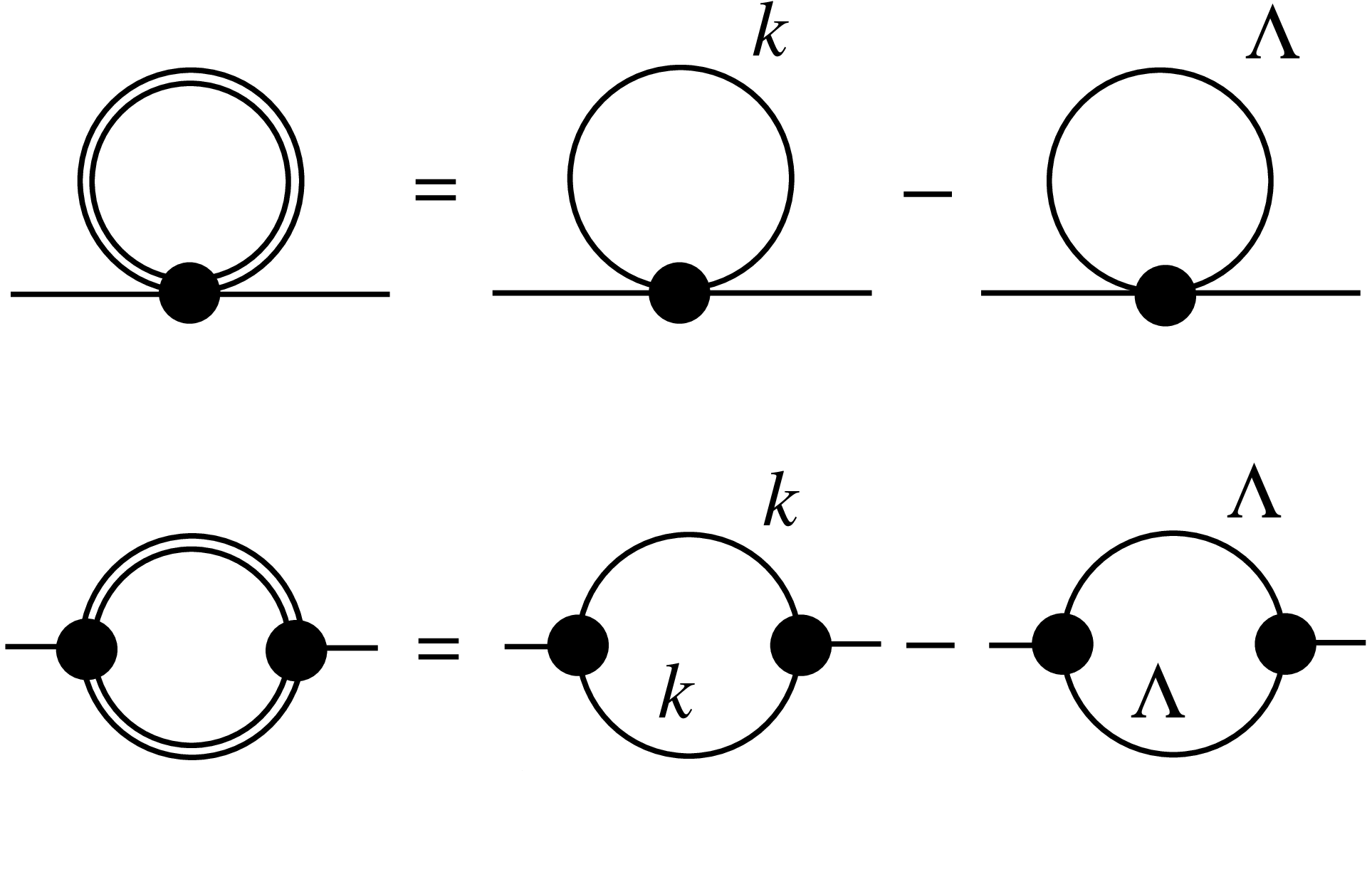,width=.5\hsize}
\end{picture}
\begin{minipage}{.8\hsize}{
{
  \small {\bf Figure 4:} Graphical representation of subtracted
  diagrams. The scale dependence of the perturbative propagator (full
  line) is due to the regulator term $R_k$; hence the index $k$ or
  $\Lambda$.}}
\end{minipage}
\end{center}
\end{figure}

The two loop contribution to the effective action is
\begin{equation}
\label{ERG2loop}
\Delta\Gamma_2 =
\s012\,\int_\Lambda^k \s0{dk'}{k'}
\ \Delta\Gamma_{1,\ pq}^{(2)}\ \partial_{t'} G_{qp}\,,
\end{equation}
where
\begin{equation}
G_{qp} =\left({1\ov S^{(2)}+R}\right)_{pq}\,.
\end{equation}
Now one uses that the only $k$-dependence of $\Delta\Gamma_1$ or its
derivatives with respect to the fields comes from the propagators $G$
within the loops. Graphically, $\partial_t G$ is given in Fig.~5.

\begin{figure}
\begin{center}
\unitlength0.001\hsize
\begin{picture}(350,100)
\put(10,35){{\Large $\partial_t$}}
\psfig{file=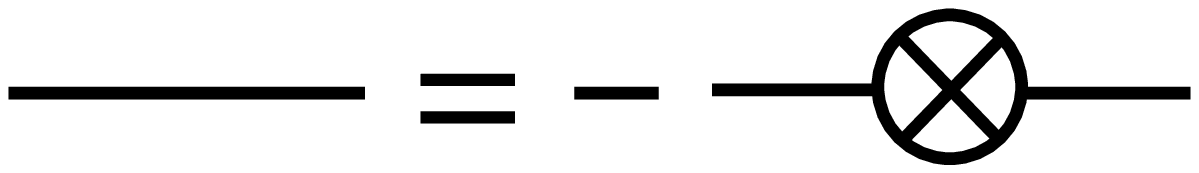,width=.4\hsize}
\end{picture}
\begin{minipage}{.8\hsize}{
{
  \small {\bf Figure 5:} Graphical representation of $\partial_t G =
  -G\,(\partial_t R)\,G$.  The $k$-dependence of $G$ is only due to the
  explicit $k$-dependence of $R_k$.}  }
\end{minipage}
\end{center}
\end{figure}

\noindent
This enables us to write \eq{ERG2loop} as a total $t$-derivative.  As
in the one loop case, for $k=0$ we approach usual perturbation theory
with the correct combinatorial factors. We get
\begin{eqnarray}
\nonumber 
\Delta\Gamma_2
\tab = \tab 
\int_\Lambda^k \s0{dk'}{k'}
\left\{ 
\s014\ 
\left(  G_{pp'}\ S_{p'pqq'}^{(4)}
      - G_{pp'}\ S_{p'rq}^{(3)}\ G_{rr'}\ S_{r'pq'}^{(3)}
\right)_\Lambda^k   
\partial_{t'} G_{q'q}
\right\}
\\ \di \nonumber  
\tab = \tab 
\int_\Lambda^k \s0{dk'}{k'}\
\s014 \partial_{t'}
\left\{\s012\, G_{pp'}\ S_{p'pqq'}^{(4)}\ G_{q'q}
      -\s013\, G_{pp'}\ S_{p'rq}^{(3)} \ G_{rr'}\ S_{r'pq'}^{(3)}\ G_{q'q}
      -{\rm subtractions}
\right\}
\\[1ex] \di 
\tab = \tab 
\left[ \s018\,    G_{pp'}\ S_{pp'qq'}^{(4)}\ G_{q'q}
      -\s01{12}\, G_{pp'}\ S_{p'qq'}^{(3)} \ G_{qr}\ S_{prr'}^{(3)}\ G_{r'q'}
\right]_{\rm ren.}, 
\label{profield} 
\end{eqnarray}
where the subscript ${}_{\rm ren.}$ indicates that these are
renormalised diagrams due to the subtractions at $\Lambda$.  Note that
the sun-set diagram in \eq{profield} is completely symmetric under
permutations of the propagators, which has lead to the factor $\s013$;
schematically written as: $(G)^2 \partial_t G= \s013 \partial_t
(G)^3$. For illustration we present in Fig.~6 the diagrams for the
term in curly brackets in the first line in \eq{profield}.  Employing
the identity displayed in Fig.~5 the expression in Fig.~6 is easily
rewritten as a total $t$-derivative. The calculation presented in
\eq{profield} is most easily carried out this way.\step

\begin{figure}
\begin{center}
\unitlength0.001\hsize
\begin{picture}(450,150)
\psfig{file=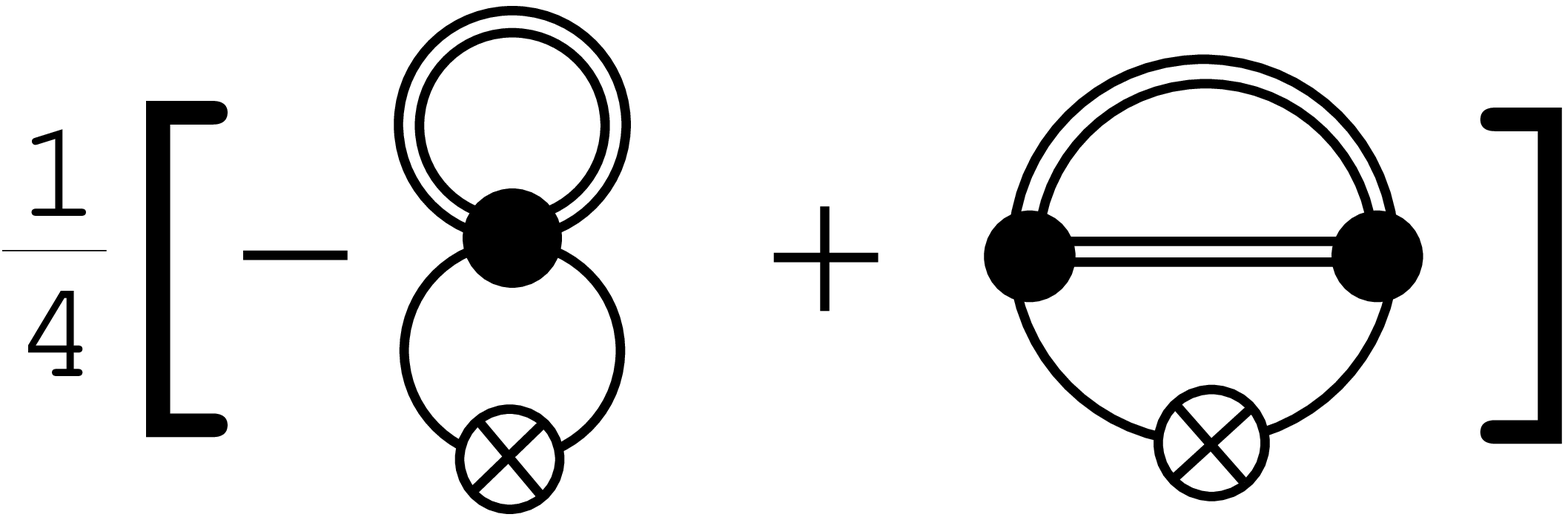,width=.45\hsize}
\end{picture}
\begin{minipage}{.65\hsize}{
{
  \small {\bf Figure 6:} The integrand in curly brackets of
  \eq{profield}, first line.}}
\end{minipage}
\end{center}
\end{figure}

\begin{figure}
\begin{center}
\unitlength0.001\hsize
\begin{picture}(500,150)
\psfig{file=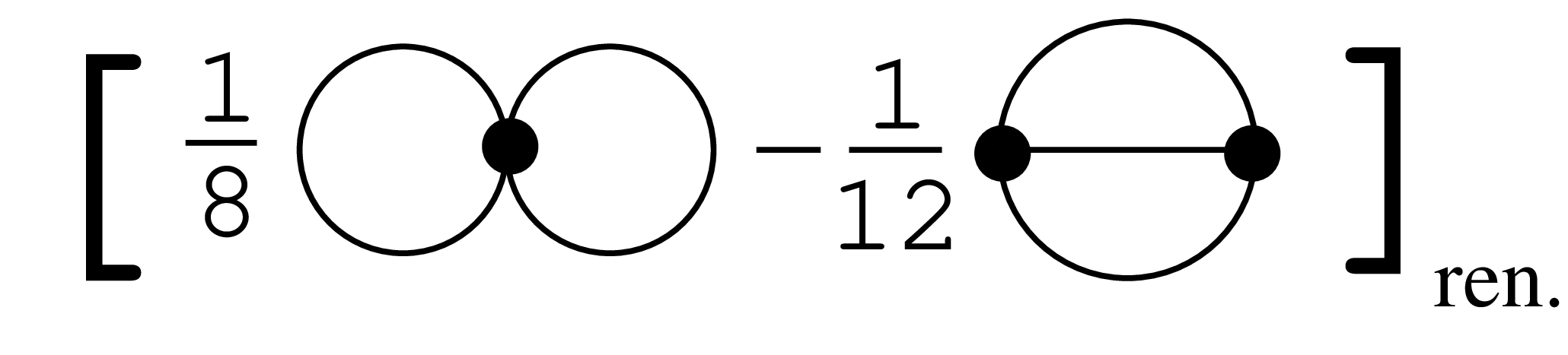,width=.5\hsize}
\end{picture}
\begin{minipage}{.85\hsize}{
{
  \small {\bf Figure 7:} two loop contribution to the effective action
  as given by \eq{profield}, last line.}}
\end{minipage}
\end{center}
\end{figure}
 
This analysis can be easily extended to any loop order. The integrands
can always be rewritten as total $t$-derivatives. Thus, the result is
independent of the regulator $R$.

\subsection{Consistent ERG flows}

Some applications of the ERG are such that a part of the quantum
fluctuations are already contained in the initial theory: in these
cases, the initial effective action is not trivial. Let us mention two
examples. First, it is possible to relax the condition \eq{3} on the
cut-off, thus starting at a point, where some (large) momentum
fluctuations are already integrated out. The control about truncations
to the starting point is very good. The neglection of power counting
irrelevant terms in the perturbative regime should only inflict
deviations of the order $(k^2/\Lambda^2)^n$ at some IR scale
$k\ll\Lambda$. Pivotal for such a picture to work is the {\it
  exactness} of the flow itself.  \step

Second, another important example are ERG flows for field theories at
finite temperature. Proposals have been put forward, which rely
on decoupling quantum fluctuations and thermal fluctuations
\cite{Litim:1998nf}. Here, the flow equation displays an integrating out
of the latter ones whereas the initial effective action contains the
quantum fluctuations. Of course, this picture only works in particular
situations where a neglection of the quantum fluctuations is feasible
or in regimes where their contributions to the effective action at
zero temperature are well under control.
\step

We conclude, that the applicability of consistent ERG flows hinges on 
their exactness. This is an important statement in view of
the applicability of other RG flows.

\section{Proper-time renormalisation group}\label{SectionPTRG}

In the remaining part of the paper we discuss one loop improved RG
flows.  In this section we consider so-called proper-time RG flows.
We show that proper-time flows in general do not reproduce the
perturbative loop expansion.  The consequences for approximations and
predictive power are discussed.

\subsection{Derivation} \label{PTRG-Derivation}

The starting point is the equation for the one loop effective action,
\begin{eqnarray}\label{1loop}
\Gamma^{\rm 1-loop}_\Lambda=S_{\rm cl}
+\s012 \Tr\ln S^{(2)}
\end{eqnarray}
The trace in \eq{1loop} is ill-defined and requires an UV
regularisation. Oleszczuk proposed an UV regularisation by means of a
Schwinger proper-time representation of the trace
\cite{Oleszczuk:1994st},
\begin{eqnarray}\label{RegDef}
\Gamma^{\rm 1-loop}_\Lambda=S_{\rm cl}
-\s012 \int\0{ds}{s}f(\Lambda,s)\Tr\exp\left(-s\,S_{\rm cl}^{(2)}\right)\,.
\end{eqnarray} 
The regulator function $f(\Lambda,s)$ provides an UV cut-off. Sending
the UV scale to $\infty$ should reduce \eq{RegDef} to the standard
Schwinger proper-time integral \cite{Schwinger:1951nm}. This happens
for the boundary condition $f(\Lambda\to\infty,s)=1$. \Eq{RegDef} can
be turned into a simple flow equation by also adding an IR scale $k$,
replacing $f(\Lambda,s)\to f_k(\Lambda,s)$. A flow equation w.r.t.~the
infra-red scale $k$ (and $t=\ln k$) has been proposed as
\cite{Liao:1996fp}
\begin{eqnarray}\label{PTRG}
\partial_t \Gamma_k[\phi]= -\s012 \int_0^\infty \0{ds}{s} \left( 
\partial_t
  f_k(\Lambda,s)\right) \Tr\exp\left(-s\Gamma^{(2)}_k\right)\,.
\end{eqnarray} 
Here, the classical action has been replaced by the scale-dependent
effective action $\Gamma_k$ on the right-hand side of \eq{PTRG}. This
is the philosophy of a one loop improvement. In \eq{PTRG} {\it only}
the explicit scale dependence due to the regulator term is considered.
There are a few conditions imposed on the proper-time regulator. The
UV behaviour remains unchanged if $\lim_{s\to 0}f_{k}(\Lambda,s)=0$.
It is required that
\begin{eqnarray}
\label{f1} 
\lim_{s\to\infty}f_{k\neq 0}(\Lambda,s)&=&0\\
\label{f3} 
\lim_{k\to \Lambda}f_{k}(\Lambda,s)&=&0\\
\label{f4} 
\lim_{\Lambda\to\infty}f_{k=0}(\Lambda,s)&=&1  
\end{eqnarray}
The condition \eq{f1} ensures that the theory is infrared regularised,
as the limit $s\to\infty$ corresponds to the limit of vanishing
momentum.  The condition \eq{f3} ensures that the the flow starts off
from the initial condition $\Gamma_\Lambda$. Finally, the condition
\eq{f4} ensures that the proper-time regularisation reduces to the
usual Schwinger proper-time regularisation for $k=0$. From now on, we
only consider regulators $f_k(\Lambda,s)$ of the form
\begin{eqnarray}\label{formoff}
f_k(\Lambda,s)=f(\Lambda^2 s)-f(k^2 s) &\quad {\rm with} &\quad 
\lim_{x\to \infty}f(x)=1 \quad {\rm and}\quad  \lim_{x\to 0}f(x)=0.
\end{eqnarray}
It is easily checked that $f_k(\Lambda,s)$ as defined in \eq{formoff} 
satisfies the conditions summarised in \eq{f1}--\eq{f4}. \step

\subsection{Completeness}\label{PTRG-Alternativ}

Next, we show that a general proper-time flow does not depend linearly
on the full propagator. We expand a general proper-time flow in the
following basis set of regulator functions $f$,
\begin{mathletters}\label{basis} 
\begin{eqnarray}
f(x;m)           & = & \0{\Gamma(m,x)}{\Gamma(m)} \\ 
\partial_t f(x;m)& = & \0{2}{\Gamma(m)} x^m  e^{-x}\,.
\end{eqnarray}
\end{mathletters}%
Here, $x=k^2 s$ and $\Gamma(m,x)=\int_0^xdt\,t^{m-1}e^{-t}$ denotes
the incomplete $\Gamma$-function.  The functions $f(x;m)$ have the
limits as demanded in \eq{formoff}. The set \eq{basis} spans the space
of all cut-offs with an IR behaviour controlled by the term $e^{-x}$
serving as a mass.  These flows cover all proper-time flows that have
been studied in the literature
\cite{Liao:1996fp,Floreanini:1995aj,Liao:1997nm,Schaefer:1999em,Meyer:2000bz,Papp:2000he,Bohr:2001gp,Bonanno:2001yp,Mazza:2001bp,Litim:2001hk,Zappala:2001nv,Meyer:2001zp,Schaefer:2001cn}.
General proper-time flows (fixed by choosing $\partial_t f$) are given
by linear combinations of the basis functions \eq{basis}.  Now we
consider the flow for a specific value of $m$. Inserting \eq{basis} in
\eq{PTRG}, we find
\begin{eqnarray}\label{insert}
\partial_t \Gamma_k= \Tr \int_0^\infty \0{ds}{s} 
  \0{(s k^2)^m}{\Gamma(m)} \exp -s\left(\Gamma^{(2)}_k+k^2\right)\,.
\end{eqnarray}
The trace in \eq{insert} can be written in terms of the (normalised)
eigenfunctions $\Psi_n$ of $\Gamma^{(2)}_k$ with
\begin{eqnarray}\label{eigen} 
\Gamma^{(2)}_k \Psi_n=\lambda_n \Psi_n\,.  
\end{eqnarray} 
This leads to 
\begin{eqnarray}\label{insert1}
\partial_t \Gamma_k= 
\sum_n \int_0^\infty \0{ds}{s} \0{(s k^2)^m}{\Gamma(m)} 
\exp -s\left(\lambda_n+k^2\right)\,.
\end{eqnarray}
For commuting the sum over $n$ and the $s$-integration we have used
that the eigenvalues obey $\lambda_n+k^2\geq 0\ \forall n$. This condition is
not a restriction as it has to hold for a well-defined proper-time flow
\eq{insert}. A similar condition also applies to ERG flows \eq{ERG}: 
$\Gamma^{(2)}+R_k\geq  0$. By
performing the $s$-integration we arrive at
\begin{eqnarray}\label{PTRG-CS}
\partial_t \Gamma_k= \sum_n\left({k^2\over
 \lambda_n+k^2}\right)^m
= \Tr 
\left({k^2\over \Gamma_k^{(2)}+k^2}\right)^m. 
\end{eqnarray}
The operator kernel inside the trace is the $m$th power of a
Callan-Symanzik kernel. Exact flows, as discussed in detail in
App.~\ref{EOL}, have a linear dependence on the full propagator.
Hence, \eq{PTRG-CS} is not exact for $m\neq 1$. Furthermore, the
{functional dependence} of \eq{PTRG-CS} on $\Gamma^{(2)}$ depends on
the regularisation. This signals that the deviation of a general
proper-time flow from an exact flow is regularisation-dependent, which
is studied next.

\subsection{Proper time flows at two loop}\label{PTRG-Completeness}
We study the deviation of integrated proper time flows from
perturbation theory at two loop. We derive relations between flows for
general $m$ and $m+n$, where $n$ is an integer and $m$ is arbitrary.
At one loop the integrated flow equation \eq{PTRG-CS} results in
\begin{eqnarray}\label{1,m}
\Delta\Gamma_{1,m}
=\int_\Lambda^k {d k'\over k'} \left.\partial_{t'} 
\Gamma_{k'}\right|_{\rm 1-loop}= 
{1\over 2 m}\Tr \left[ \left(\frac{k'^2}{
\Gamma_{k'}^{(2)}}\right)^m
\,\,{}_2 F_1\left(m,m;m+1;-
\0{k'^2}{\Gamma_{k'}^{(2)}}\right)\right]_\Lambda^k,  
\end{eqnarray}
where ${}_p F_q(x,y;z;w)$ 
is the generalised hypergeometric series. For integer $m$, 
${}_p F_q$ can be summed up and there is a simpler representation 
\begin{eqnarray}\label{del1,m}
\Delta\Gamma_{1,m}=\int_\Lambda^k {d k'\over k'} 
\left.\partial_{t'} \Gamma_{k'}\right|_{\rm 1-loop}= 
{1\over 2}\Tr \left[ \ln \left(\Gamma_{k'}^{(2)}+{k'}^2\right)
-\sum_{n=1}^{m-1} {1\over n} 
\left({{k'}^2\over \Gamma_{k'}^{(2)}+{k'}^2}\right)^n\right]_\Lambda^k. 
\end{eqnarray}
For $k\to 0$ both formulas reproduce the one loop effective action
$\s012[\Tr\ln(\Gamma_{k}^{(2)}+{k}^2)]_{\rm ren}$. For $k\neq 0$ we
also have additional terms as opposed to the one loop integral of an
ERG flow, cf.~\eq{ERG1loop}. These terms are $m$-dependent. For
general $m$ the difference between $\Delta \Gamma_{1,m}$ and $\Delta
\Gamma_{1,m-1}$ is given by
\begin{eqnarray}\label{id2txt}
\Delta \Gamma_{1,m}-\Delta \Gamma_{1,m-1}=-\s0{1}{2(m-1)}
\left[\Tr (G\, {k'}^2 )^{m-1}\right]_\Lambda^k\,,
\end{eqnarray}
with $G=(S^{(2)}+k^2)^{-1}$. The right-hand side vanishes for $k\to
0$.  At two loop, we can relate flows with $m$ and $m'=m+n$, where $n$
is integer. The details are given in App.~\ref{secrecursive}. The key
result is the recursive relation
\begin{eqnarray}
\Delta\Gamma_{2,m}-\Delta\Gamma_{2,m-1}=
\012 \int_\infty^0 
\0{dk}{k}\,\Tr \left[({G\,{k}^2})^{m-1}\, G\,  
\,{\delta^2\over(\delta\phi)^2}\Tr\, (G \,{k}^2)^{m-1} G\,S^{(2)}
\right]\,,
\label{recursivetxt} \end{eqnarray}
apart from irrelevant terms from the different renormalisation
procedures for the two flows. A similar relation was already presented
in \cite{Litim:2001ky}. It is connected to \eq{recursivetxt} by a
partial integration, see App.~\ref{secrecursive}. Using
\eq{recursivetxt} recursively, we find
\begin{eqnarray}
\Delta\Gamma_{2,m}=\Delta \Gamma_{2,m-n}+
\012\sum_{l=m-n}^{m-1}\int_\infty^0 
\0{dk}{k}\, \Tr \left[({G\,{k}^2})^{l}\, G\,
\,{\delta^2\over(\delta\phi)^2}\Tr\, (G \,{k}^2)^{l} G\,S^{(2)}
\right]\,.
\label{m-ntxt} \end{eqnarray}
The difference \eq{m-ntxt} depends on arbitrarily high powers of the
fields and does not vanish. \Eq{m-ntxt} provides a constructive proof
that proper-time flows, in general, are non-exact. Let us assume for a
moment that the proper-time flow for a particular $m_0$ is exact. Then
it follows from \eq{m-ntxt} that {\it all} flows with $m=m_0+n$ for
integer $n$ are {\it not} exact, because the corresponding terms
\eq{m-ntxt} do not vanish identically in the fields. This has an
immediate consequence for flows with integer $m$: The Callan-Symanzik
flow ($m=1$) is exact.  Therefore any flow with integer $m>1$, or any
linear combination thereof, is not exact.  \step

Let us close with two comments. We have found regulator dependent
terms at two loop. Hence, the proper time flow \eq{PTRG} does not
represent a total $t$-derivative. One could think that the proper-time
flow \eq{PTRG} is improved by also taking into account the
$t$-derivative of $\Gamma^{(2)}_k$,
\begin{eqnarray}
\partial_t \Gamma_k= -\s012 \int_0^\infty \0{ds}{s} 
\Tr\left( \partial_t
  f_k(\Lambda,s) - sf_k(\Lambda,s)\,
  \partial_t\Gamma^{(2)}_k \right)
\exp\left(-s\Gamma^{(2)}_k\right) \,.  \label{ImprovedPTRG}
\end{eqnarray} 
The flow equation \eq{ImprovedPTRG} is, in contrast to \eq{PTRG}, a
total $k$-derivative. Its end point does not depend on the
regularisation. However, the end point is the functional $\Gamma$
which solves $\Gamma=S_{\rm cl}+ \Tr\ln\Gamma^{(2)}|_{\rm ren}$. This
equation is not satisfied by the full effective action.\step

A second comment concerns another extention of proper-time flows,
discussed in App.~\ref{SectionCS}. Consider the flow
\begin{eqnarray}\label{relatetxt}
\partial_t \Gamma_k- 
\sum_{n=1}^{m-1} F_{n,m}\partial_t^{n+1} \Gamma_k = 
\Tr \left({k^2\over \Gamma_k^{(2)}+k^2}\right)^m  +
\Tr\, F_k[\partial_t\Gamma_k^{(2)},...,\partial^{m-1}_t\Gamma_k^{(2)};
\Gamma_k^{(2)}]\,. 
\end{eqnarray} 
The coefficients $F_{n,m}$ and $F_k$ are defined in
App.~\ref{SectionCS}. The flow \eq{relatetxt} is exact and $\Gamma_k$
obeys the usual Callan-Symanzik equation. The first term in
\eq{relatetxt} is the standard proper time flow \eq{PTRG-CS}. The new
terms in \eq{relatetxt} are proportional to the flow of
$\Gamma_k^{(2)}$ and to higher order scale derivatives of $\Gamma_k$.

\subsection{Consistency}\label{PTRG-Consistency} 

We have shown that proper-time flows are not complete.  We are left
with the question whether proper-time flows are consistent.  In this
case the initial effective action $\Gamma_\Lambda$ is non-trivial and
must be known explicitly. Let us first argue that {\it any} flow
trivially represents an exact flow by the following construction.  The
initial effective action $\Gamma_\Lambda$ is given as a function of
the flow and the full effective action $\Gamma_0$ by
\begin{eqnarray}\label{bs}
\Gamma_\Lambda[\Gamma_0] = 
\Gamma_0 + \int_0^\Lambda {d k\over k} \partial_{t} \Gamma_{k}[\Gamma_0]. 
\end{eqnarray} 
At least within a loop expansion this is possible, as $n$-loop order
contribution to $\Gamma_\Lambda$ depend on the flow to loop order
$n-1$.  The only condition for the global construction is the
existence of flow trajectories from the full effective action
$\Gamma_0$ to $\Gamma_\Lambda$. For such a scenario to be applicable,
the initial effective action \eq{bs} has to be known explicitly. Then
the flow is consistent. If the initial condition is not known
explicitly, the flow cannot be integrated. This consideration implies
that proper-time flows are not consistent: the flow is not complete,
and we don't have any further information about $\Gamma_\Lambda$,
except the trivial one encoded in \eq{bs}.  This observation makes it
interesting to investigate possible enhancements of proper-time flows,
which is done in Sect.~\ref{SectionCCRG}.

\begin{figure}
\begin{center}
\unitlength0.001\hsize
\begin{picture}(250,180)
\psfig{file=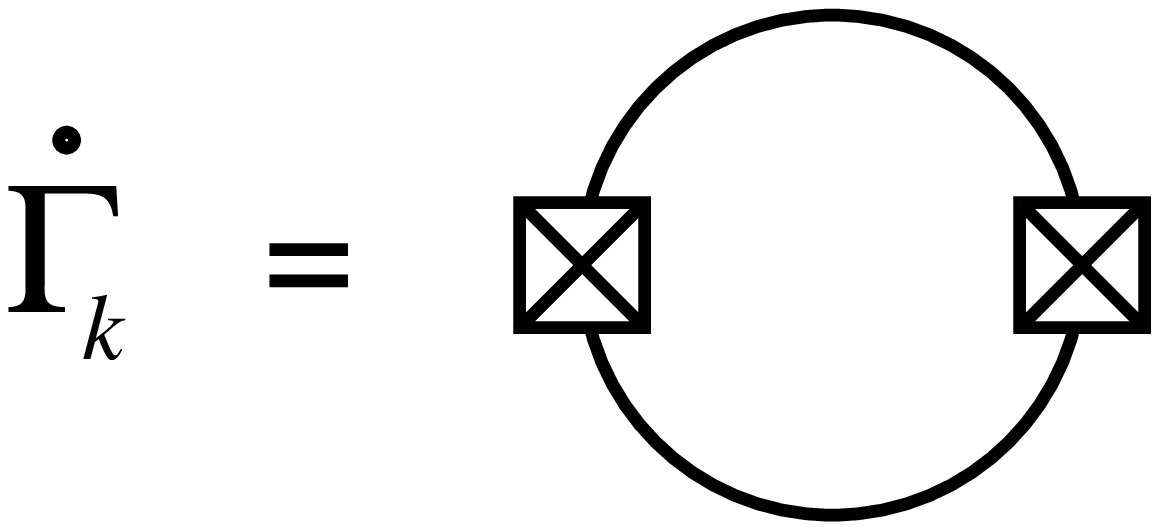,width=.25\hsize}
\end{picture}
\hskip.5\hsize
\begin{minipage}{.85\hsize}{
    {\small {\bf Figure 8:} Graphical representation for the
      proper-time RG equation \eq{m=2}. The proper-time flow
      \eq{PTRG-CS}, for integer $m$, corresponds to a loop with $m$
      propagator lines and $m$ insertions $k^2$. It reduces to the CS
      flow \eq{CS} for $m=1$.}}
\end{minipage}
\end{center}
\end{figure}

\subsection{Example}\label{m=2example}

We illustrate our findings with a simple example by considering $m=2$.
A short account of this calculation was already given in
\cite{Litim:2001ky}. In the condensed notation introduced in
Sect.~\ref{ERG_Completeness}, the proper-time flow with $m=2$ is
\begin{eqnarray}\label{m=2} 
\partial_t \Gamma_k = \left({ k^4\ov {(\Gamma_k}^{(2)}+k^2)^2}
\right)_{qq}\,, 
\end{eqnarray} 
where the kernel is the square of a Callan-Symanzik kernel, and $q$
denotes momenta.  The flow is depicted in Fig.~8.  The line in Fig.~8
stands for the full field-dependent propagator
$(\Gamma^{(2)}+k^2)^{-1}$, the crossed square stands for the insertion
$k^2$.  This has to be compared with the ERG flow in Fig.~1. \step

The one loop contribution from the integrated flow \eq{m=2} can be
read off from \eq{del1,m} as
\begin{equation}\label{PT1loop}
  \Delta\Gamma_1= \s012 \left[\ln \left( k^2+S^{(2)}\right)\right]_{qq}-
\s012 k^2 G_{qq}
\end{equation}
with
\begin{equation}\label{G}
G_{qq'}= \left({ 1\ov S^{(2)}+k^2}\right)_{qq'}. 
\end{equation}
The two loop contribution is
\begin{eqnarray}\label{2-loop} 
\Delta\Gamma_2 = -2 \int_\Lambda^k \s0{dk'}{k'} 
\ \Delta\Gamma_{1,\ qq'}^{(2)}\ (G\ k'^2\ G\ k'^2\ G)_{q'q}. 
\end{eqnarray} 
In \eq{2-loop}, it is understood that $G$ \eq{G} depends on
$k'$ under the integral. From \eq{PT1loop}, we obtain
\begin{eqnarray}
\Delta\Gamma_{1,\ qq'}^{(2)}&= &\nonumber
\s012 
\left[  G_{pp'}
\left(  S_{p'pqq'}^{(4)}
      - S_{p'rq}^{(3)}\ G_{rr'}\ S_{r'pq'}^{(3)} \right)
\right.\\
&&
\quad\left.+ (G\ k'^2\, G)_{pp'}
\left(   S_{p'pqq'}^{(4)}
      -2 S_{p'rq}^{(3)}\ G_{rr'}\ S_{r'pq'}^{(3)} \right)
\right]_\Lambda^k\,.
\label{DG1} 
\end{eqnarray}
Notice the difference to \eq{profield1}. Graphically, \eq{DG1} is
given in Fig.~9, where we resort to the definitions in Fig.~4. Lines
represent field dependent perturbative propagators, vertices represent
field dependent classical vertices.

\begin{figure}
\begin{center}
\unitlength0.001\hsize
\begin{picture}(700,170)
\psfig{file=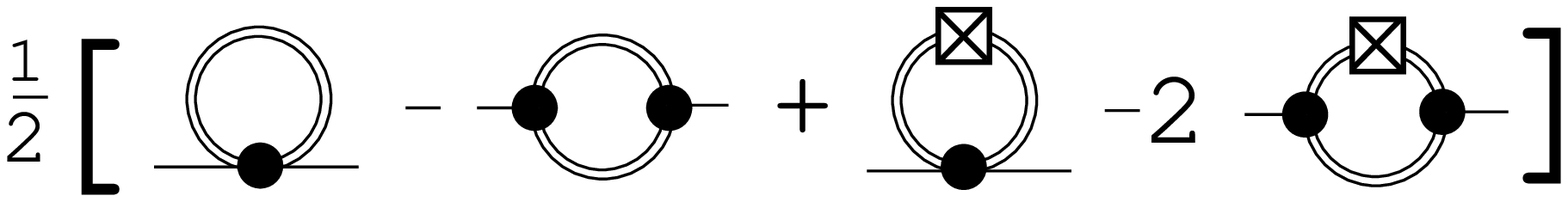,width=.75\hsize}
\end{picture}
\begin{minipage}{.8\hsize}{
    {\small {\bf Figure 9:} The one loop correction to the propagator
      $\Delta\Gamma_{1}^{(2)}$ for the specific flow \eq{m=2}. Notice
      the two additional terms which appear in comparison to the ERG
      flow, cf.~Fig.~3.}}
\end{minipage}
\end{center}
\end{figure}

In comparison to the ERG result for the one loop propagator in Fig.~3
there are two additional diagrams in Fig.~9.  Inserting \eq{DG1} into
\eq{2-loop}, we end up with
\begin{eqnarray}\nonumber 
\lefteqn{\Delta\Gamma_2  = \int_\Lambda^k \s0{dk'}{k'}
\left\{
-\left[ 
G_{pp'}\left(  S_{p'pqq'}^{(4)}
            - S_{p'rq}^{(3)}\ G_{rr'}\ S_{r'pq'}^{(3)}\right)
\right]_\Lambda^k \
(G\ k'^2\, G\ k'^2\, G)_{q'q}
\right.}
\hspace{2.7cm}
\\ \di 
\tab \tab  
\left. -
\left[ 
(G\ k'^2\, G)_{pp'}
\left(   S_{p'pqq'}^{(4)}
      -2 S_{p'rq}^{(3)}\ G_{rr'}\ S_{r'pq'}^{(3)}
\right)
\right]_\Lambda^k
(G\ k'^2\, G\ k'^2\, G)_{q'q}
\right\}\,. 
\label{PTRG2loop}
\end{eqnarray}
The integrand in \eq{PTRG2loop} has the graphical representation given
in Fig.~10. 
\step

\begin{figure}
\begin{center}
\vskip-1cm
\unitlength0.001\hsize
\begin{picture}(600,250)
\psfig{file=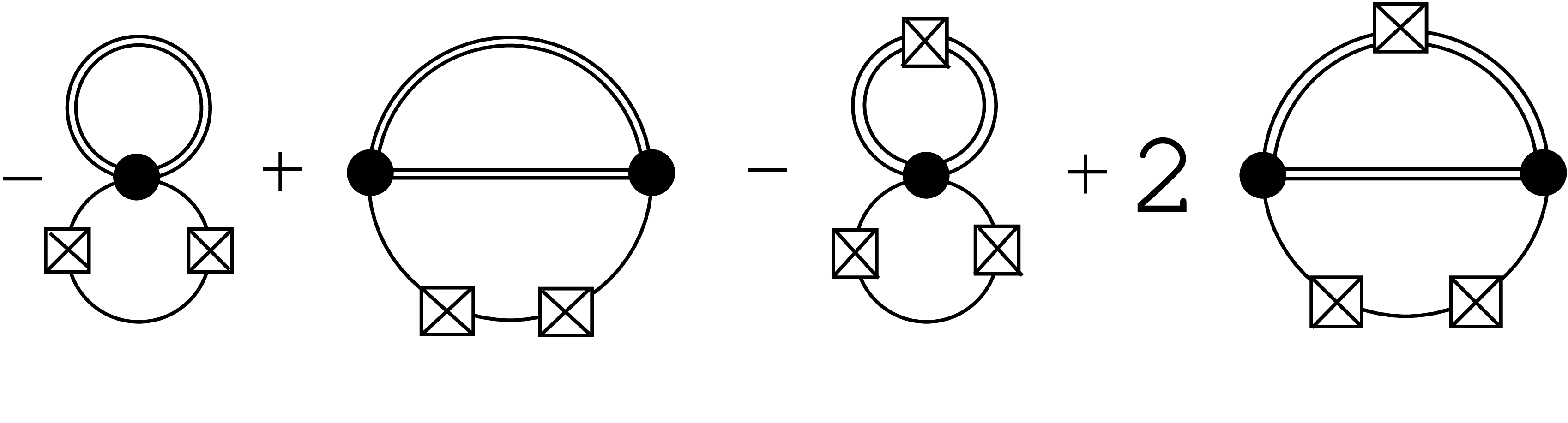,width=.65\hsize}
\end{picture}
\begin{minipage}{.8\hsize}{
    {\small {\bf Figure 10:} The integrand in \eq{PTRG2loop}.  See
      \eq{CSfinal} and Fig.~15 for comparison.}}
\end{minipage}
\end{center}
\end{figure}

Next, we compare our findings with a generalised Callan-Symanzik flow
\eq{2der} discussed in App.~\ref{CS-Example}. This flow is exact.
It differs from the proper-time flow \eq{m=2} only by loop terms
proportional to the flow $\partial_t\Gamma_k^{(2)}$. Graphically, the
difference between the flows is given by the second term in Fig.~14.
At two loop, we compare the integrands as given in Fig.~10 and
Fig.~15, respectively.  The first two terms in Fig.~15 and Fig.~10
agree whereas the last two terms are different. More specifically, the
last two terms in Fig.~15 have $G\, k^2\, G$ as the bottom line,
whereas we have $G\, k^2\, G\, k^2\, G$ in Fig.~10.  It is this
difference which makes it impossible to rewrite the integrand in
\eq{PTRG2loop} as a total derivative. \step

\begin{figure}
\begin{center}
\unitlength0.001\hsize
\begin{picture}(350,270)
\psfig{file=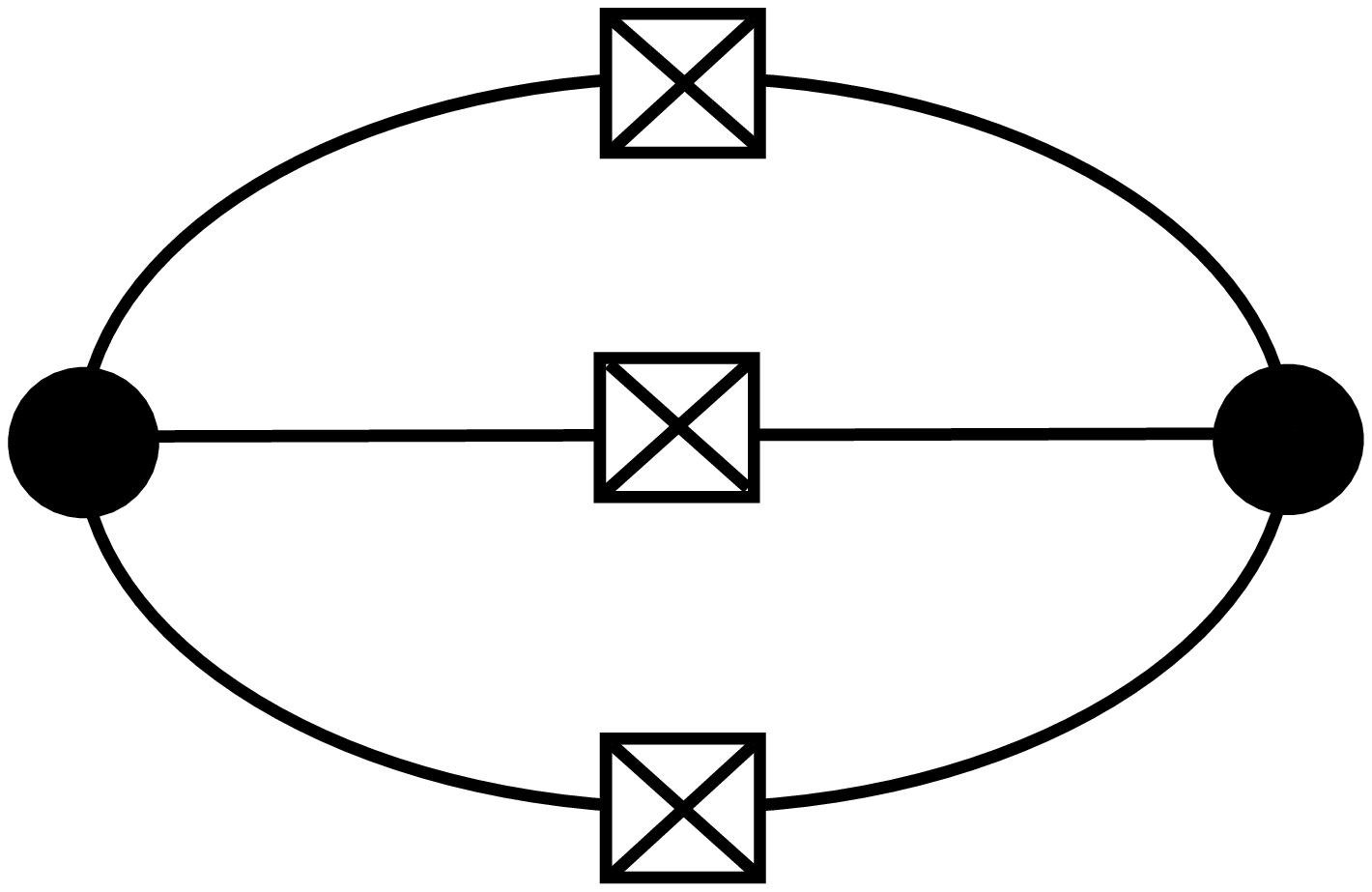,width=.32\hsize}
\end{picture}
\begin{minipage}{.8\hsize}{
{
  \small {\bf Figure 11:} The non-standard term in \eq{deviation}. See
  also \eq{nonstandard}.}}
\end{minipage}
\end{center}
\end{figure}

Now, let us expand \eq{PTRG2loop} about the correct two loop result
\eq{profield}. After some algebra, we arrive at
\begin{eqnarray}
\Delta\Gamma_2 &=&
\left[
 \s018\,      G_{pp'} \ S_{p'pqq'}^{(4)}\ G_{q'q}
-\s0{1}{12}\, G_{pp'} \ S_{p'rq}^{(3)} \ G_{rr'}\ S_{r'pq'}^{(3)} 
 {G}_{q'q}\right]_{\rm ren.}
\nonumber\\ &&  
-\s012 \, \int_\Lambda^k \s0{dk'}{k'} \
(G\ k'^2\, G)_{pp'} \ S_{p'rq}^{(3)}\
(G\ k'^2\, G)_{rr'} \ S_{r'pq'}^{(3)}\
(G\ k'^2\, G)_{q'q} \,. 
\label{deviation}
\end{eqnarray}
A simple consistency check on \eq{deviation} is to take its derivative
w.r.t. $k$. This leads to the kernel of \eq{PTRG2loop}. The first line
in \eq{deviation} corresponds to the correct two loop result. The
second line denotes the deviation from standard perturbation theory.
The integrand in the second line of \eq{deviation} is the non-standard
diagram depicted in Fig.~11.  The second term on the right-hand side
of \eq{deviation} is the term on the right-hand side of the recursive
relation \eq{recursivetxt} for $m=2$ [see also App.~\ref{CS-Example},
Eqs.~\eq{difference} and \eq{nonstandard}].  The last term on the
right-hand side of \eq{deviation} cannot be absorbed in
renormalisation constants. It contains arbitrary powers in fields and
momenta and does not integrate to zero in the limit $k\to0$ and
$\Lambda\to\infty$. For massive theories both limits are safe.  This
term displays a non-trivial deviation of the present proper-time flow
from perturbation theory. The form of the integrand is that of the
sunset graph where all propagators have been substituted by their
squares.  \step

To be more explicit, consider a massive $\phi^4$-theory with mass $M$
and quartic interaction $\s01{4!} \lambda \int d^d x \phi^4$. The
contribution of the non-standard diagram to the propagator is obtained
after taking the second derivative with respect to the fields in
\eq{deviation} at $\phi=0$. We find
\begin{eqnarray}\label{propphi4} 
\lambda^2 \int_\infty^0 \s0{d k}{ k} \int \frac{d^d q}{(2\pi)^d}
\frac{d^d l}{(2\pi)^d}\,  
{k^2\over (k^2+M^2+q^2)^2} 
{k^2\over (k^2+M^2+l^2)^2} 
{k^2\over (k^2+M^2+(l+q-p)^2)^2} 
\end{eqnarray}
The integrand it strictly positive. Hence the integral is
non-vanishing.  Moreover it has a non-trivial momentum dependence.
This can be seen by evaluating the limits $p\to 0$ and $p\to\infty$.
For $p\to 0$ we are left with a non-vanishing constant. For
$p\to\infty$ the expression in \eq{propphi4} vanishes.

\section{Multiplicative regularisation}\label{SectionOther}

In this section we discuss a recent suggestion for a one loop improved
RG \cite{Liao:2000yu}, which is based on an operator regularisation of
the one loop effective action. The starting point of
\cite{Liao:2000yu} is the regularised form of the one loop effective
action,
\begin{eqnarray}\label{sb0}
\Gamma^{\rm 1-loop}_k=\s012\Tr \left(\rho\,\ln {S}^{(2)}\right)\,.
\end{eqnarray}
Here, $\rho$ provides a regularisation of the otherwise ill-defined
trace in \eq{sb0}. In the limit $k\to 0$ the regularisation is removed
and $\rho\to 1$.  Taking the $t=\ln k$ derivative of \eq{sb0} and
using the condensed notation introduced in \eq{notation} leads to
\begin{eqnarray}\label{sb01}
\partial_t \Gamma^{\rm 1-loop}_k=\s012\left(\ln {S}^{(2)}\right)_{qq'}
\partial_t
\rho_{q'q}.
\end{eqnarray}
Again one resorts to the idea of a one loop improvement and
substitutes ${S}^{(2)}$ on the right-hand side of \eq{sb01} with
$\Gamma_k^{(2)}$.  This leads to the final form of the one loop
improved flow,
\begin{eqnarray}\label{sb}
\partial_t \Gamma_k=
\s012\left(\ln \Gamma_k^{(2)}\right)_{qq'}
\partial_t
\rho_{q'q}\,.
\end{eqnarray}
The factorisation of the regulator $\rho$ makes numerical as well as
analytical calculations easily accessible. In Ref.~\cite{Liao:2000yu},
the flow \eq{sb} has been studied to leading order in the derivative
expansion. As the flow \eq{sb} depends on the logarithm of
$\Gamma^{(2)}$, it cannot be exact. \step

We would like to understand the structure of the deviation more
explicitly and compute the two loop effective action. The one loop
effective action is 
\begin{eqnarray}\label{sb1loop}
\Delta\Gamma_1=
\int_\Lambda^k \s0{dk'}{k'}
\left. \partial_{t'} \Gamma_{k'}\right|_{\rm 1-loop}
=\left(
\s012 \left(\ln S^{(2)}\right)_{qq'}\rho_{q'q}\right)_{\Lambda}^k.
\end{eqnarray}
The two loop effective action is
\begin{eqnarray}\label{sb2loop}
\Delta\Gamma_2=\s012 \int_\Lambda^k \s0{dk'}{k'}
 \ \Delta{\Gamma^{(2)}_{1,\ qq'}} \ G_{q'p}\ \partial_{t'} \rho_{pq}\,,
\end{eqnarray}
where $G=1/S^{(2)}$.  We rewrite the expression on the right-hand side
of \eq{sb2loop} as a total derivative using that the only
$k$-dependence of $\Delta \Gamma_1^{(2)}$ is given by $\rho$.  We
finally get
\begin{eqnarray}\nonumber
\Delta\Gamma_2\tab = \tab
\int_\Lambda^k \s0{dk'}{k'}
\left(  \s014 G_{pp'}\, \rho_{p'r}\, S_{rpqq'}^{(4)}
      - \s014 G_{pp'}\,\rho_{p'r}\, S_{rr'q}^{(3)}\,G_{r's}\,
S_{spq'}^{(3)}
\right)_\Lambda^k G_{q's'}\, \partial_{t'}\rho_{s'q}
\\ \di
\tab = \tab
\left[ \s018 (\rho\, G)_{pp'}\, S_{p'pqq'}^{(4)}\, (\rho\, G)_{q'q}
      -\s018 (\rho\, G)_{pp'}\, S_{p'rq}^{(3)}
             G_{rr'}\, S_{r'pq'}^{(3)}\, (\rho\, G)_{q'q}\right]_{\rm
ren.}.
\label{profieldsc}
\end{eqnarray}
For $k=0$, the two loop result \eq{profieldsc} is independent of the
regularisation. The integrand in \eq{profieldsc} has the graphical
representation given in Fig.~12. Fig.~13 shows the two loop
contribution of the flow \eq{sb}, corresponding to the last line in
\eq{profieldsc} at $k=0$.

\begin{figure}
\begin{center}
\unitlength0.001\hsize
\begin{picture}(500,150)
\epsfig{file=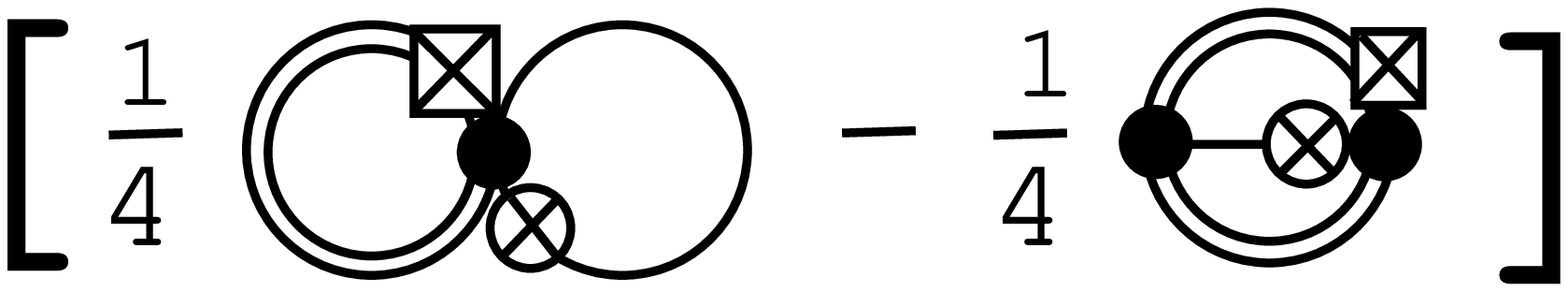,width=.5\hsize}
\end{picture}
\begin{minipage}{.8\hsize}{
    \small {\bf Figure 12:} The integrand of \eq{profieldsc}, first
    line. Notice that the insertions $\rho$ and $\partial_t\rho$ are
    always attached to a vertex.}
\end{minipage}
\end{center}
\end{figure}

\begin{figure}
\begin{center}
\unitlength0.001\hsize
\begin{picture}(500,150)
\epsfig{file=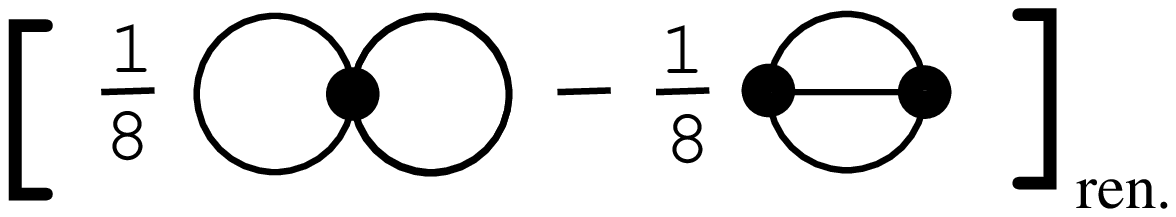,width=.5\hsize}
\end{picture}
\begin{minipage}{.8\hsize}{
    \small {\bf Figure 13:} The two loop effective action derived from
    \eq{sb}, and as given by the last line of \eq{profieldsc}.}
\end{minipage}
\end{center}
\end{figure}

The combinatorial factor for the sunset graph is not the correct one.
How does this come about? In the ERG case, one deals with expressions
which are, qualitatively, of the form $(G)^n\partial_t G= {1\ov
  n+1}\partial_t (G)^{n+1}$. Stated differently, {\it all} propagators
are regularised.  In the RG equation \eq{sb}, there is {\it one}
regulator insertion $\rho$ for {\it each} loop, regardless, how many
propagator are contained in the loop. The first diagram in Fig.~12
contains two loops and two propagators, leading to the correct
combinatorial factor in Fig.~13. The sunset diagram contains two loops
but three propagators, therefore, the combinatorial factor comes out
too big by $\s032$. \step

To sum up, in contrast to ERG flows which are based upon a
regularisation of the full inverse propagator, the one loop improved
flow \eq{sb} is based on a regularisation of the {\it logarithm} of
the full inverse propagator. This choice has been motivated in order
to facilitate computations, and to find simple expressions for the
flow. As it turns out, it is precisely this form of the regularisation
which is ultimately responsible for the mismatch with standard
perturbation theory beyond one loop.

\section{Exact proper-time flows}\label{SectionCCRG}
In this section we relate proper-time flows to exact flows, both,
within given approximations and as closed formal expressions.

\subsection{Proper-time representation of ERG flows}
\label{ERG-PTRG}

We have already introduced a representation of proper-time flows which
is quite close to the ERG (see Sect.~\ref{PTRG-Alternativ}). Let us
now investigate a proper-time representation of ERG equations. This
allows us to study the map from ERG to proper-time flows more directly
as done in \cite{Litim:2001hk}. We start with the ERG \eq{ERG} which
can be rewritten as
\begin{eqnarray}\label{flowjan}
\partial_t \Gamma_k 
={1\over 2} \Tr\, 
\partial_t R_k\,\int_0^\infty ds\, \exp -s(\Gamma^{(2)}_k+R_k).
\end{eqnarray}
It is easy to see that the flow equation \eq{flowjan} is well-defined
in both the ultra-violet and the infra-red.  We now turn 
\eq{flowjan} into
\begin{eqnarray}\label{PTERG}
\partial_t \Gamma_k ={1\over 2}
\int_0^\infty {ds \over s}\  \Tr 
\left(F_k[sR_k;s\Gamma^{(2)}_k]
\,\exp{-s\Gamma^{(2)}_k}\right)\,, 
\end{eqnarray}
in order to facilitate the comparison of ERG flows to proper-time
flows \eq{PTRG}. Here, the operator $F_k[A,B]$ is given as
\begin{eqnarray} \label{F}
F_k[A;B]&=& (\partial_t A)
\exp(-A+K[-(A+B),B])
\\[1ex]
\label{DK}
K[A,B]&=& \sum_{n=1}^\infty {(-)^n\ov n+1}
\sum_{p_i+q_i\geq 1} {1\ov 1+\sum_{i=1}^n p_i} \prod_{i=1}^n{
(\ad\, A)^{p_i}\ov 
p_i!}{(\ad\,B)^{q_i} \ov q_i !}[A], 
\end{eqnarray}
where $\ad B[A]=[B,A]$ and $(\ad B)^0[A]=A$.  Equations \eq{F} and
\eq{DK} can be deduced from the Baker-Campbell-Hausdorff-formula:
\begin{eqnarray}\label{BCH}
e^{A+B+K[A,B]} e^{-B}=e^{ A}\,.  
\end{eqnarray} 
The term $K[-s(\Gamma^{(2)}_k+R),sR]$ vanishes for
$[\Gamma^{(2)}_k,\,R]=0$.  Now we compare the representation
\eq{PTERG} of ERG flows with proper-time flows \eq{PTRG}. We already
know that proper-time flows and ERG flows are not equivalent.
Comparing the kernels, this information is encoded in
\begin{eqnarray}\label{strict} 
\partial_t  f_k(\Lambda,s)
\neq
F_k[sR_k;s\Gamma^{(2)}_k]\,,
\end{eqnarray} 
which states that no field- and momentum-independent function $f$ can
be found to match the right-hand side of \eq{strict}.  Indeed, the
right-hand side carries physical information about the theory due to
$\Gamma^{(2)}_k[\phi]$.  \step

Still, there are two options left to overcome \eq{strict}. First, the
expressions in \eq{strict} are integration kernels. Within given
approximations, the integrals could agree despite of the kernels being
qualitatively different. This possibility is worked out in
Sect.~\ref{PTRG-PP}.  Second, one may consider generalised proper time
regulators, by allowing for an additional dependence on
$\Gamma^{(2)}_k$. This is addressed in Sect.~\ref{genPTRG}.

\subsection{Derivative expansion}\label{PTRG-PP}


Next, we study ERG and proper-time flows to lowest order in a
derivative expansion, where wave function renormalisations are not
present. Here, we explicitly map regulators $R_k\rightarrow f_k[R_k]$. 
The inverse map does not exist in general. 
In \cite{Litim:2001hk}, a similar analysis was performed on the level
of the threshold functions. The effective action in this approximation is 
\begin{eqnarray}\label{0deriv} 
\Gamma_k[\phi]=\int d^d x 
\big[\partial_\mu\phi\partial_\mu\phi+U_k(\phi)\big] 
\end{eqnarray}
and, consequently,
\begin{eqnarray}\label{0deriv1} 
\Gamma^{(2)}_k[\phi](p^2)= p^2+U''_k(\phi).
\end{eqnarray}
The scale dependent part of the effective action is the potential
$U_k$. We only consider constant fields $\phi_0$ in the flow.  In
this approximation, we have
\begin{eqnarray}\label{vanish}
[\Gamma^{(2)}_k[\phi_0],\,R]=0
\end{eqnarray}
which implies that $F_k$ in \eq{PTERG} depends only on $R_k$. Then we
cast the ERG equation in a proper-time form, also using \eq{F} 
\begin{eqnarray}\nonumber
\partial_t \Gamma_k& =&
\displaystyle 
\s012 \Tr 
\int_0^\infty {ds \over s}\, s\,
(\partial_t R_k)\, \exp( -s R_k)\exp-s(p^2+U_k'')\\ 
&=&
\displaystyle 
\s014\Omega_d \int_0^\infty {ds \over s}
\left(s \,
\int_0^\infty {dy\over  y} y^{d/2} (\partial_t R_k) \exp-s (R_k(y)+ y) 
\right) \exp-s\, U''_k\,,   
\label{flowjanderive}\end{eqnarray}
where $\Omega_d$ is the volume of the $d$-sphere over $(2\pi)^d$,
$\Omega_d =2 ((2 \pi)^{d/2}\Gamma(d/2))^{-1}$ and $y=q^2$.  This has
to be compared with \eq{PTRG} in this approximation. After performing
the momentum integration in \eq{PTRG} we get
\begin{eqnarray}\label{PTRGp0}
\partial_t \Gamma_k= \s014\Omega_d \int_0^\infty {ds \over s}\left(s^{-d/2}
\,\partial_t f_k(\Lambda,s)\right)\exp{-s\, U''_k}. 
\end{eqnarray}
\Es{flowjanderive} and \eq{PTRGp0} are identical for the following choice 
of $f_k$:  
\begin{eqnarray}\label{approx}
\partial_t f_k(\Lambda,s)= -{s^{1+d/2}\over \Gamma(d/2)} 
\int_0^\infty {dy \over y} y^{d/2} (\partial_t R_k) \exp{-s (R_k+ y)}, 
\end{eqnarray} 
\Eq{approx} defines a map $R_k\to f_k(R)$. Thus it is guaranteed that
there is always a function $f_k$ corresponding to a choice of $R$.
Next, we show that the opposite is not the case. \step

\Eq{approx} fixes the behaviour of $f_k(R)$ for $s\to 0$, which is the
UV-limit and $s\to\infty$, which is the IR-limit. We restrict
ourselves to regulators with constant IR-limit: $R(x\to 0)\propto
k^2$. Moreover we demand that
\begin{eqnarray}\label{no-fluc}
\min_{y}\left(y+R(y)\right)= c_0 k^2 \quad {\rm with} \quad c_0>0. 
\end{eqnarray}
\Eq{no-fluc} implies that we have an IR regularisation. If we would
take $c_0\leq 0$ we introduce poles in the momentum integration of the
ERG. Thus, \eq{no-fluc} leads to an exclusion of wildly fluctuating
regulators $R$. With \eq{no-fluc} we deduce the following limit
behaviour of $f_k(R,s)$:
\begin{eqnarray}\label{0}
\lim_{s\to 0}|\partial_t f_k(R,s)|\tab <\tab  
s^{d/2+1}\exp({-s c_0 k^2})\, C[R]\\\di 
\lim_{s\to \infty}|\partial_t f_k(R,s)|\tab < \tab s^{d/2+1}
\exp({-s c_0 k^2})\,C[R]\,,
\label{infty}\end{eqnarray}
where 
\begin{eqnarray}\label{R}
C[R]={1\over \Gamma(d/2)}
\int_0^\infty {dy\over y^{1-d/2}} \,\partial_t R(y).
\end{eqnarray} 
and the exponential factor in \eq{0} is sub-leading and has only been
introduced for symmetry reasons. These limits only make sense for
$C[R]<\infty$ (no infra-red divergent cut-offs).  Infra-red divergent
cut-offs including the sharp cut-off are even more severely limited in
the infra-red for $s\to \infty$. Only if $f_k$ obeys both limits
\eq{0} and \eq{infty}, the corresponding regulator $R_k$ exists.
Here, the relevant limit is $s\to\infty$. \step

It is left to investigate the r$\hat{\rm o}$le of the constant $c_0$.
We assume to have found a regulator $f_k(R,s)$ which precisely matches
the boundary value of the IR limit: $f_{\rm ext.}  (R,s\to
\infty)=c_f\, s^{d/2+1} \exp({-c_0 s k^2})$. The UV behaviour is
irrelevant for the integration of the flow.  The normalisation $c_f$
follows from the conditions \eq{f1} -- \eq{f4}, leading to
\begin{eqnarray}\label{normf} 
f_{\rm ext}(R,s)= 
\0{2(c_0 s k^2)^{d/2+1}}{\Gamma(d/2+1)} \exp({-c_0 s k^2}).
\end{eqnarray}
Since \eq{normf} depends only on the product $c_0 k^2$, we can
reabsorb $c_0$ in the infrared scale and set it to one, $c_0=1$.
\step

Next we verify some of the explicit examples given earlier in
\cite{Litim:2001hk}.  We insert several cut-offs into the right-hand
side of \eq{approx} ($x=s k^2$) to find the proper-time analogues. For
the optimised regulator \cite{Litim:2001up}, the sharp cut off and the
mass-like regulator
\begin{eqnarray}\label{optver}
R^{\rm opt}_k(q^2)  &=& (k^2-q^2) \theta(k^2-q^2)\,, \\
R^{\rm sharp}_k(q^2)&=& \lim_{c\to \infty} c\, \theta(k^2-q^2) \,, \\
R^{\rm mass}_k(q^2) &=& k^2\,, 
\end{eqnarray} 
we find the proper-time analogues as
\begin{eqnarray}
\tab \partial_t f_k(\Lambda,s)&= 
-\04d\01{\Gamma(d/2)} x^{d/2+1} \exp{-x}\,,\\ 
\tab \partial_t f_k(\Lambda,s)&= 
{2\over \Gamma(d/2)} x^{d/2} \exp{-x}\,,\\ 
\tab \partial_t f_k(\Lambda,s)&= 
-x \exp{-x}. 
\end{eqnarray} 
The optimal cut-off \cite{Litim:2001up} precisely matches both limits
\eq{0} and \eq{infty} (for $c_0=1$).  In this sense it is an extremum
of the allowed space of $f_k$.  \step

In summary, there is only a narrow
window for proper-time regulators $f_k$ which are images of ERG
regulators $R$. We find that regulators $f_k(R,s)$ are generally given by 
\begin{eqnarray}\label{window}
\partial_t f_k(R,s)=
\int_{d/2}^{d/2+1} d m\, \0{2 x^m b(m) }{\Gamma(m)} \exp(-x)
\ \,{\rm with}\,\,
\int_{d/2}^{d/2+1} d m\, b(m)=1.
\end{eqnarray}
For other proper-time regulators there is no corresponding ERG
regulator $R$. The upper boundary $m_{\rm max}=d/2+1$ follows from the
IR limit \eq{infty}. The lower boundary $m_{\rm min}=d/2$ is the
demand of UV finiteness. It can be relaxed to $m_{\rm min}=1$, thus
including Callan-Symanzik flows as a boundary.  \step

\subsection{Generalised proper-time flows and background fields}
\label{genPTRG}
Finally we derive a {\it generalised} proper-time flow which is both
consistent and complete.  Since \eq{strict} cannot be satisfied, we
seek for a convenient generalisation of the proper-time regulator. As
we cannot get rid of the operator dependence on the right-hand side of
\eq{strict} we have to allow for field- and momentum-dependent
functions $\partial_t f_k(\Lambda,s)$.  A key property of a
proper-time flow \eq{PTRG} is that the operator trace only depends on
the operator $\Gamma_k^{(2)}$. Maintaining this simple structure, and
allowing for a field- and momentum-dependent regulator leads to
\begin{equation}\label{general-f}
\partial_t
f_k(\Lambda,s)\to \partial_t f_k[\Lambda,s;\Gamma_k^{(2)}]\,.
\end{equation}
Such a generalised proper-time flow is equivalent to an ERG flow, if
\begin{eqnarray}\label{restrict} 
\partial_t  f_k[\Lambda,s;\Gamma_k^{(2)}]
\stackrel{!}{=}
F_k[sR_k;s\Gamma^{(2)}_k]\,.
\end{eqnarray}
In order to satisfy \eq{restrict}, the regulator $R_k$ must depend
solely on $\Gamma_k^{(2)}$ and its $t$-derivative,
\begin{equation}\label{general-R}
R_k(q^2)\to R_k[\Gamma_k^{(2)}]\,.
\end{equation}
In order not to spoil the one loop structure of the ERG flow equation,
$R_k$ cannot depend on the {\it full} propagating field. The only
admissible dependence of $R_k$ on $\Gamma^{(2)}_k$ comes via
background fields. For details of a background field formulation of
the ERG (for gauge theories) we refer the reader to
\cite{Reuter:1994kw,Litim:1998qi,Freire:2000bq}. Here we mention the
important facts by restricting ourselves to a scalar theory: in the
background field formalism, the full field $\phi=\bar\phi+\varphi$ is
split into the background field $\bar\phi$ and the fluctuation field
$\varphi$. The effective action depends on the fields $\phi$ and 
$\bar\phi$, $\Gamma_k=\Gamma_k[\phi,\bar\phi]$. As the propagating field is
$\varphi$, the regulator $R_k$ can only depend on
$\Gamma_k^{(2)}[\bar\phi,\bar\phi]$, where
$\Gamma_k^{(2)}[\phi,\bar\phi]
:=\delta^2\Gamma_k[\phi,\bar\phi]/{(\delta\phi)^2}$. The cut-off
term depends on $\bar\phi$ and it follows that
$\Gamma_k[\phi,\bar\phi]\neq \Gamma_k[\phi]$ for $k\neq 0$. Finally
such a flow depends also on $\partial_t
\Gamma_k^{(2)}[\bar\phi,\bar\phi]$.  For the explicit form of the flow
notice that the operator $K[A,B]$ in \eq{F} vanishes for $[A,B]=0$.
Hence, a vanishing commutator
\begin{equation}\label{[GR]=0}
[\Gamma^{(2)}_k,\,R_k]=0
\end{equation}
implies that the operator $F_k$ in \eq{PTERG} becomes under the trace 
\begin{equation}\label{PTgeneralised}
F_k[s R_k]= \left( s\partial_t R_k\right) \exp(-sR_k)= -2 s \left(\Gamma_k^{(2)}R'-R-\s012 
\partial_t \Gamma_k^{(2)} R'\right) \exp(-sR_k)
\end{equation}
In this case, the representation \eq{PTERG} simplifies tremendously.
\Eq{[GR]=0} holds trivially at $\bar\phi=\phi$, where $R_k$ is a
function of $\Gamma^{(2)}_k[\bar\phi,\bar\phi]$.  The flow is
\begin{eqnarray}\label{genRG}
\partial_t \Gamma_k[\phi,\phi]= \s012 \int_0^\infty \0{ds}{s}\  
\Tr\ F_k\Big[s R_k[\Gamma^{(2)}_k[\phi,\phi]]\Big]\, 
\exp\left(-s\Gamma^{(2)}_k [\phi,\phi]\right)
\end{eqnarray} 
with $F_k$ given by \eq{PTgeneralised}. The corresponding ERG flow is
given by
\begin{eqnarray}\label{genRG1}
\partial_t \Gamma_k[\phi,\phi]= \s012  
\Tr\ {1\over\Gamma_k^{(2)}[\phi,\phi]
+ R_k\Big[\Gamma_k^{(2)}[\phi,\phi]\Big]}
\ \partial_t R_k\Big[\Gamma_k^{(2)}[\phi,\phi]\Big]
\end{eqnarray}
In summary, the following picture has emerged: we have defined a
generalised proper-time flow for an effective action based on the
background field formalism. It differs from the standard one by terms
proportional to $\partial_t\Gamma^{(2)}$. These terms make the flow
consistent and complete. It can be mapped to an ERG flow at vanishing
fluctuation fields. The flow equation is not closed, because it
depends on $\Gamma_k^{(2)}[\phi,\phi]$. The output of the flow
equation is $\Gamma_k[\phi,\phi]$ and does not entail the information
for $\Gamma_k^{(2)}[\phi,\phi]$, which requires the derivative
w.r.t.~the first argument. The background field dependence is
controlled by a separate equation \cite{Litim:1998qi,Freire:2000bq}.
The flow \eq{genRG1}, apart from being an interesting subject by its
own right, gives a clear definition on the limits of proper-time
flows.\step

\section{Discussion}\label{SectionDiscussion}

We have studied the completeness and consistency of different one loop
RG flows. We summarise the main results and their implications.\step

Consistency and completeness of a flows are directly related to the
propagator dependence of the flow, which, for an exact flow, has to be
linear. The linearity is important for a recursive perturbative
integration of the flow. For exact flows, the integrated flow at a
given loop order contains the same diagrams with identical
combinatorial factors as standard perturbation theory.  ERG flows at
two loop served as an illustration of these facts. \step

For proper-time flows, we have shown that they are not complete. This
result is based, first, on a structural analysis of the proper-time
flow. When written in the form \eq{PTRG-CS}, it is apparent that the
functional dependence of the flow on the full propagator is non-linear
--- except when it matches the Callan-Symanzik flow. Second, we have
formally integrated the flow up to two loop order. As a result, we
have explicitly established that the integrated proper-time flow
deviates from perturbation theory. The deviation of fully integrated
proper-time flows (when the cut-off is removed) from fully integrated
exact flows turns out to be regulator-dependent.  Proper-time flows
are also not consistent, because it is not known beforehand which part
of perturbation theory is missing along the flow. \step

An analogous analysis has been applied to the one loop improved flow
\eq{sb}. We found that \eq{sb} is neither complete nor consistent, for
arbitrary regulator. The main structural reason for this fact is that
the flow depends logarithmically on the full propagator for {\it any}
regulator, and not linearly. This structure entails that, first, the
perturbative loop expansion does not lead to the correct combinatorial
factors, and, second, that the deviation from perturbation theory is
independent on the regulator. This last property is in marked contrast
to proper-time flows. There, we have seen that the functional
dependence of the flow on $\Gamma^{(2)}$ is regulator dependent, as
is, consequently, the deviation from perturbation theory.\step 

Links between proper time flows and exact flows have been discussed in
 Sect.~\ref{SectionCCRG}. This enabled us to provide information about
 the inherent approximation they represent to exact flows. We
 established links between exact flows and standard proper-time flows
 along three different lines:

First, we provided an explicit equation for the deviation of proper
time flows from Callan-Symanzik flows. This deviation is given by the
difference between \eq{relate} and \eq{PTRG-CS}.  Essentially,
proper-time flows lack additional contributions from two sources.
There are additional one loop terms proportional to scale derivatives
of $\Gamma_k^{(2)}$, and a sum of higher scale derivatives of
$\Gamma_k$.

Second, it is possible to relate proper-time flows to exact
flows within specific approximations.  To leading order in the
derivative expansion, we derived explicit maps from ERG flows to
proper-time flows and discussed their properties. It has also been
shown that higher orders of the derivative expansion cannot be mapped
onto ERG flows. 

Third, we constructed {\it generalised} proper-time flows \eq{genRG}.
These flows can be mapped in a closed form to specific ERG flows,
which established both, completeness and consistency for \eq{genRG}.
Similar to the generalised Callan-Symanzik flow, they differ from the
standard proper-time flow only through higher order terms proportional
to the flow of $\Gamma_k^{(2)}$. This philosophy, however, applies
only within a background field method. \step

These results have important implications.  Most notably, they make
the intrinsic approximation of a proper-time flow explicit. This makes
it possible to link approximations to proper-time flow to
approximations to the full theory and allows to discuss predictive
power within the formalism. For its applications, it is important to
know how results based on standard proper-time flows are affected by
the additional terms. For example, for $3d$ scalar theories at
criticality, a particular proper-time flow
\cite{Litim:2001hk,Mazza:2001bp} has lead to critical exponents, which
agree remarkably well with experiment or Monte Carlo simulations. From
the present analysis, it emerged that the underlying exact flow
contains additional contributions already to leading order in a
derivative expansion. These terms are expected to modify the physical
predictions, and it remains to be seen whether these corrections are
quantitatively large or small. We hope to report on this issue in
near future. \step\step\step

{\bf Acknowledgments:}
DFL thanks the Institute for Theoretical Physics III, University of
Erlangen, and JMP thanks CERN for hospitality and financial support.
The work of DFL has been supported by the European Community through
the Marie-Curie fellowship HPMF-CT-1999-00404.\step

\setcounter{section}{0}
\renewcommand{\thesection}{\Alph{section}}
\renewcommand{\thesubsection}{\arabic{subsection}}
\renewcommand{\theequation}{\Alph{section}.\arabic{equation}}

\section{Structure of one loop exact flows}\label{EOL}

In this paper, we have discussed renormalisation group flows whose
striking feature is their one loop nature. It is precisely this
property which facilitates numerical implementations, as we need not
to cope with overlapping integrals. In this appendix, we derive the
most general form of one loop flows that are {\it exact}. We consider  
one loop flows with the general form
\begin{eqnarray}\label{oneloopexact}
k\partial_k \Gamma_k[\phi]=
\Tr \ f_k[\Gamma_k^{(2)}],
\end{eqnarray}
where $f_k[\Gamma_k^{(2)}](p,q)$ is a smooth function of its arguments.
It depends both explicitly and implicitly, via $\Gamma_k^{(2)}$, on
momenta. We demand that $\Gamma_{k=0}$ is the full quantum effective
action.  The structure of flows given by \eq{oneloopexact} covers all
flows discussed in the literature and in the present work. Note that
$f_k$ may also have some intrinsic dependence on running couplings and
vertices of the theory. Trivially there are no overlapping momentum
integrals in \eq{oneloopexact}.\step

As it stands, a flow of the form \eq{oneloopexact} can be derived
within a one loop improvement philosophy. Then, $f_k$ just encodes the
information of the cut-off procedure at one loop. We want to know,
what restrictions are posed upon $f_k$ if we demand that
\eq{oneloopexact} is an {\it exact} flow, {\it i.e.}\ a flow which has
a first principle derivation, say from a path integral representation
of the theory. The path we take, is the following. First, we derive
the most general form of flows for the functional $Z$. Then we discuss
convenient parameterisations of such flows.  Finally we translate our
findings to flows of the effective action $\Gamma_k$ via a Legendre
transform.\step

Let us consider the functional $Z[S,J]$. The first argument of $Z$
indicates the classical action, about which the theory is quantised. A
general flow of $Z[S,J]$ can be described by the flow of an operator
insertion ${\cal O}_k$ depending on a cut-off scale $k$. We define
\begin{eqnarray}
Z[S,J;{\cal O}_k]= \int d\phi\,{\cal O}_k[\phi] \exp\left(-S[\phi]
+\int \phi\,J\right).
\label{intW}
\end{eqnarray}
In particular, we demand $\lim_{k\to 0} {\cal O}_k[\phi]=1$. In this
limit, \eq{intW} reduces to $Z[S,J;1]\equiv Z[S,J]$, the full
generating functional. The flow of $Z[S,J;{\cal O}_k]$ is given by
\begin{eqnarray}
k\partial_k
Z[S,J;{\cal O}_k]=
 \int d\phi\ k\partial_k{\cal O}_k[\phi]\ \exp\left(-S[\phi]
+\int \phi\,J\right)\label{genW}
\end{eqnarray}
Thus, a general flow of $Z$ is just given by the expectation value
$\langle k\partial_k {\cal O}_k[\phi]\rangle_{S,J}$.  However,
expectation values of $\phi^n$ with $n>2$ involve multi-loop
contributions in the full propagator.  This can be seen as follows: We
expand ${\cal O}_k[\phi]$ in powers of $\phi$. Terms in the expansion
have the form
\begin{eqnarray*}
\left\langle \int_{p_1,...,p_n}
k\partial_k {\cal O}_k^{(n)}
(p_1,...p_n)\prod_{i=1}^n \phi(p_i) \right\rangle_{S,J}.
\end{eqnarray*}
This expectation value can be written in terms of the Schwinger
functional $W[S,J]= \ln Z[S,J]$ as
\begin{eqnarray}\nonumber
\lefteqn{\int d\phi\, \int_{p_1,...,p_n}
 {\cal O}_k^{(n)}(p_1,...p_n) \prod_{i=1}^n \phi(p_i) \exp\left(-S[\phi]
+\int \phi\,J\right)}\hspace{6cm}\\
&=& \int_{p_1,...,p_n}
 {\cal O}_k^{(n)}(p_1,...p_n) \prod_{i=1}^n{\delta\over \delta J(p_i)}
\exp W[S,J].
\label{phi^n}
\end{eqnarray}
Thus it depends on all functional derivatives $\delta^i
W/(\delta J)^i$ with $i\leq n$. Next we check how \eq{phi^n} is
expressed in terms of the full propagator
$(\Gamma^{(2)})^{-1}=\delta^2W/(\delta J)^2$. The
propagator enters in the recursive relation
\begin{eqnarray}\label{recursiveapp}
\prod_{i=1}^n{\delta\over \delta J(p_i)} W[S,J]=
\int_q \left(\frac{1}{\Gamma^{(2)}}(p_1,q)
{\delta\over \phi(q)}\right)
\prod_{i=2}^{n}{\delta\over \delta J(p_i)} W[S,J=\s0{
\delta\Gamma}{\delta\phi}].
\end{eqnarray}
Consequently {\it any} expectation value \eq{phi^n}, expressed in
terms of $\Gamma_k$ and its derivatives, for $n>2$ contains multi-loop
terms.  This leads to the first important result: flows, 
which are exact already at one loop can only involve expectation
values of at most two fields.\step

However, the argument above did not make use of the form of the
classical action $S$ entering the exponent in the path integral. We
can always use a redefinition of $S$ as follows
\begin{eqnarray}
k\partial_k {\cal O}_k[\phi]
\exp\left(-S[\phi]\right) =
\int_{p_1,p_2}\phi(p_1)\hat O_k(p_1,p_2) \phi(p_2)\exp
-\left(S[\phi]+\tilde{\cal O}_k[\phi]\right)
\label{redefineS}
\end{eqnarray}
where $\tilde{\cal O}_k$ depends on the choice of $\hat{\cal O}_k$ and
${\cal O}_k$. Take ERG flows as an example. Here ${\cal O}_k=\exp\,
\s012\int \phi R\phi$. Choosing $\hat{\cal O}_k=\s012 
k\partial_k\, R$ we have $\tilde{\cal O}_k=\s012\int \phi
R\phi$. Note that in general $\tilde{\cal O}_k$ is highly non-local.
We conclude that general flows can be written as one loop exact flow
with
\begin{eqnarray}
k\partial_k
Z[S+\tilde{\cal O}_k,J;\hat{\cal O}_k]=
\int d\phi\, \int_{p_1,p_2} \phi(p_1)
\hat{\cal O}_k(p_1,p_2)\phi(p_2)\, \exp
\left(-S+{\tilde{\cal O}_k}+\int \phi\,J\right).
\label{genOneloop}
\end{eqnarray}
Our findings can be summarised in the following statement: Any flow --
if represented as a one loop exact flow for $\Gamma_k$ of the form
\eq{oneloopexact} -- depends linearly on the full propagator. In
consequence, the most general form for the function $f_k$ is
\begin{eqnarray}
f_k[\Gamma^{(2)}](p_1,p_2)=
\int_{q} \hat{\cal O}_k(p_1,q) {1\over \Gamma^{(2)}}(q,p_2)\,,
\end{eqnarray}
where $\Gamma_k$ is the Legendre transform of $\ln W[S+\tilde{\cal
  O}_k,J]$.  Finally, we mention that only those functions
$\tilde{\cal O}$, which are polynomial in the fields, have simple
properties for $k\to \infty$. Furthermore, the functional $\Gamma_k$
matches simple boundary conditions only if $\tilde{\cal O}$ is
quadratic in the fields. These requirements are met for ERG flows.

\section{Generalised Callan-Symanzik flows}\label{SectionCS}

In this appendix, we discuss RG flows based on a mass term $R=k^2$.
The resulting flow is a Callan-Symanzik (CS) flow
\cite{Callan:1970yg}. This flow can be brought into the more standard
form of the Callan-Symanzik equation in case we had introduced
anomalous dimensions. On the basis of the CS flow we construct flows which are
similar in form to proper-time flows.  We restrict ourselves to the 
discussion of massive theories, in order to avoid some particular
problems with massless ones. Massless theories can be dealt with as
well, but the additional problems there are of no relevance for our
purposes.\step

Employing the notation introduced earlier, the CS flow is simply given by 
\begin{eqnarray}\label{CS} 
\partial_t\Gamma_k = \Tr {k^2\over \Gamma_k^{(2)}+k^2}\,.
\end{eqnarray} 
We stress that the CS flow is not precisely an ERG flow as defined
above since it fails to satisfy condition \eq{3}. In particular, the
CS flow does not admit the Wilsonian interpretation of the flow: in
contrast to the ERG case, at every fixed scale $k$, the momentum
integration is not regularised in the UV and all momenta contribute to
the flow. There is the necessity of an additional UV renormalisation
of the flow, not required for the ERG. This problem has been discussed
in detail in \cite{Litim:1998nf}. For the present purposes we can
neglect this intricacy. \step

The integrated CS flow \eq{CS} gives the full quantum effective
action. Let us now address a slightly different flow, given by the 
difference of the CS flow and the flow of the CS flow,
\begin{eqnarray}\label{close}
(\partial_t-\s012 \partial_t^2)\Gamma_k = 
(1-\s012\partial_t)\Tr {k^2\over \Gamma_k^{(2)}+k^2}=
\Tr\left({k^2\over \Gamma_k^{(2)}+k^2}\right)^2 +
\s012\Tr {k^2\over (\Gamma_k^{(2)}+k^2)^2}\partial_t\Gamma_k^{(2)}.
\end{eqnarray} 
\Eq{close} represents a flow for $(1-\s012 \partial_t)\Gamma_k$. Such
a flow trivially escapes the linearity constraint on general one loop
exact flows derived in appendix~\ref{EOL}. It involves higher
derivatives of a general one loop exact flow with respect to $k$. This
is signalled by the term proportional to $\partial_t\Gamma_k^{(2)}$
on the right hand side. Consequently it does not match the allowed structure 
on the right hand side of \eq{oneloopexact}. Note, however, that
$\Gamma_k$ satisfies the CS-equation in agreement with 
appendix~\ref{EOL}.  Integrating the flow displayed in \eq{close}
leads to the effective action. For $k\to 0$, we arrive at
\begin{eqnarray}\label{integrate}
\Gamma_\Lambda-\s012 \left.\partial_t\Gamma_k
\right|_{k=\Lambda}
+ \int_\Lambda^0 \frac{dk}{k} 
(\partial_t-\s012 \partial_t^2)\Gamma_k = \Gamma_0 
-\s012 \left[\Tr {k^2\over \Gamma_k^{(2)}+k^2}\right]_{k=0}= \Gamma_0. 
\end{eqnarray} 
The initial condition for such the flow \eq{close} is
$\Gamma_\Lambda-\s012 \left.\partial_t\Gamma_k\right|_{k=\Lambda}$
which tends to the classical action for $\Lambda\to\infty$. Such a
flow is complete. However, we emphasise, that the right hand side of 
\eq{close} depends on $\Gamma_k^{(2)}$ and $\partial_t\Gamma_k^{(2)}$ and 
$\Gamma_k$ obeys the CS-equation. This is important for the iterative 
calculations done in appendix~\ref{CS-Example}. \step

This example can be extended to arbitrary
sums of derivatives $(\partial_t+\sum_n c_n \partial_t^n)\Gamma_k$.
Integrals of these flows always result in the effective action due to
the first term $\partial_t\Gamma_k$. This can be used to define the
following flow:
\begin{eqnarray}\label{relate}
\partial_t \Gamma_k- 
\sum_{n=1}^{m-1} F_{n,m}\partial_t^{n+1} \Gamma_k = 
\Tr \left({k^2\over \Gamma_k^{(2)}+k^2}\right)^m  +
\Tr\, F_k[\partial_t\Gamma_k^{(2)},...,\partial^{m-1}_t\Gamma_k^{(2)};
\Gamma_k^{(2)}]\,. 
\end{eqnarray} 
Here, $F_k[0,...,0;x]\equiv 0$ and $F_{n,m}=-\s012
\sum_{i=n}^{m-1}\s0{1}{i} F_{n-1,i}$ for $n>1$ and $F_{1,m}=\s012
\sum_{i=1}^{m-1}\s0{1}{i}$. $F_k$ is given by the terms proportional
to $\partial_t^i \Gamma_k^{(2)}$ with $i=1,...,m-1$ contained in
$\sum_{n=1}^{m-1} F_{n,m}\partial_t^{n+1} \Gamma_k$.  By construction,
the flow \eq{relate} is an exact flow. Again, as for the integrated
flow \eq{close} (see \eq{integrate}), the integral of \eq{relate} is
the full effective action:
\begin{eqnarray}\nonumber 
\lefteqn{\Gamma_\Lambda-
\sum_{n=1}^{m-1} \left.F_{n,m}\partial_t^{n} \Gamma_k\right|_{k=\Lambda}-
\sum_{n=1}^{m-1} F_{n,m}\int_\Lambda^0 \frac{dk}{k} 
\partial_t^{n+1} \Gamma_k}\hspace{5cm}\\
&=& \Gamma_0-\sum_{n=1}^{m-1} \left.F_{n,m}\ \partial_t^{n}\
\Tr \ {k^2\over \Gamma_k^{(2)}+k^2}  
\right|_{k=0}=\Gamma_0.
\label{genintegrate}\end{eqnarray} 
The integrated flow is the full effective action, as the additional
terms are proportional to powers of $k^2$. Moreover, the initial
effective action tends to the classical action for $\Lambda\to\infty$,
subject to a properly chosen renormalisation. We have shown in
Sect.~\ref{SectionPTRG} that the first term of \eq{relate} represents
a generic proper-time flow \cite{Litim:2001ky}. Hence, $\Tr\,
F_k[\partial_t\Gamma_k^{(2)},...,\partial^{n-1}_t\Gamma_k^{(2)};
\Gamma_k^{(2)}]+\sum F_{n,m}\partial_t^{n+1}\Gamma_k\neq 0$ represents 
the unavoidable deviation of a proper-time flow from an exact flow.
\step

\section{Example}\label{CS-Example}
In this appendix we calculate the two loop contribution of 
the generalised CS flow as introduced in 
App.~\ref{SectionCS} for $m=2$. This serves as a reference point 
for the proper-time flow with $m=2$, discussed in Sect.~\ref{m=2example}. 
The line of reasoning is analogous to the
one presented in Sect.~\ref{ERG_Completeness}. In the condensed
notation introduced there, the flow \eq{close} is given by
\begin{eqnarray}\label{2der}
\left(\partial_t-\s012 \partial_t^2\right)\Gamma_k = 
\left({ k^4\ov (\Gamma_k^{(2)}+k^2)^2}\right)_{qq}
+\s012 \left({k^2\ov (\Gamma_k^{(2)}+k^2)^2}\right)_{qq'}
\partial_t\Gamma^{(2)}_{k,\ q'q}. 
\end{eqnarray}
The right-hand side of \eq{2der} has the graphical representation
given in Fig.~14.

\begin{figure}
\begin{center}
\unitlength0.001\hsize
\begin{picture}(500,420)
\epsfig{file=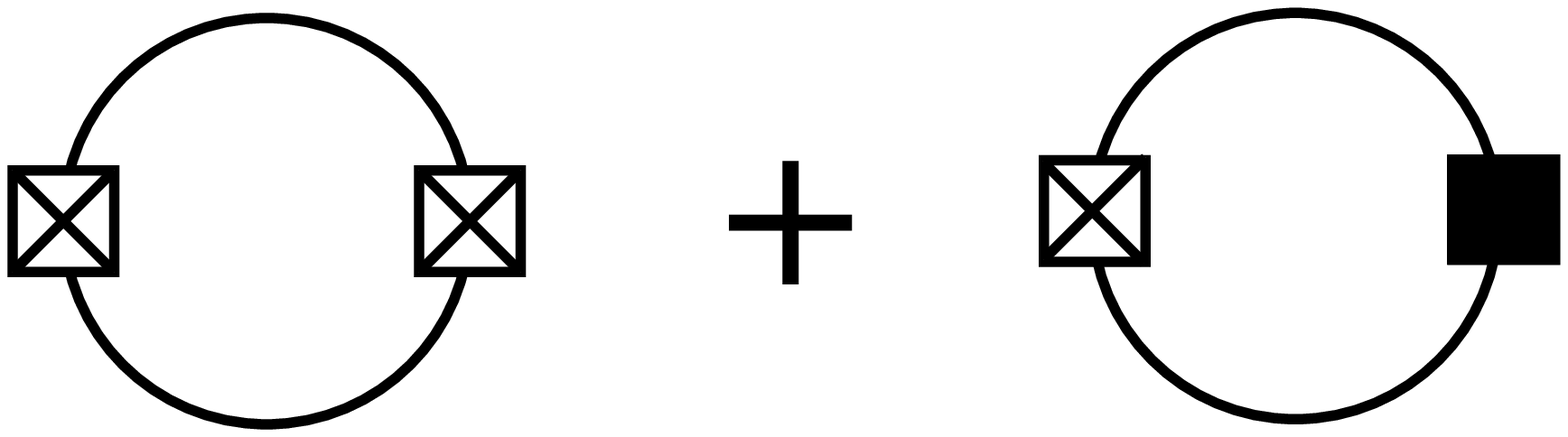,width=.5\hsize}
\end{picture}
\hskip.5\hsize
\vskip-4cm
\begin{minipage}{.7\hsize}{
    {\small {\bf Figure 14:} Graphical representation of \eq{2der}.
      The black box denotes the insertion
      $\s012\partial_t\Gamma^{(2)}_k$. The second term corresponds to
      the difference w.r.t.~the proper-time flow \eq{m=2}, given in
      Fig.~8.}}
\end{minipage}
\end{center}
\end{figure}

Expanding \eq{2der} in loop orders we arrive at
\begin{eqnarray}
\Delta\hat\Gamma_2 &=& 
\int_\Lambda^k \s0{dk'}{k'}
\left.
\left(\partial_{t'}-\s012 \partial_{t'}^2 \right)
\Gamma_{k'}\right|_{\rm 2-loop} 
\nonumber\\ \label{CS2-loop} 
&=&
\int_\Lambda^k \s0{dk'}{k'} 
\left( - 2\Delta\Gamma_{1,\ pq}^{(2)} 
  (G\ k'^2\, G\ k'^2\, G)_{qp}
+ \s012 (G\ k'^2\, G)_{pq}\ \partial_{t'} \Gamma_{k',\ qp}^{(2)}\right). 
\label{condenseCS}\end{eqnarray}
The hat in $\Delta\hat\Gamma_2$ indicates that $\Delta\hat\Gamma_2$
has a diagrammatic expansions different from $\Delta\Gamma_2$. Note
also that on the right-hand side $\Delta\Gamma_1$ is the one of the CS
flow \eq{CS}. Now we use that
\begin{eqnarray}
\Delta\Gamma_{1,\ qq'}^{(2)} &=&  
\s012\, 
\left[ G_{pp'}
\left(  S_{p'p qq'}^{(4)}
      - S_{p'r q}^{(3)}\ G_{rr'}\ S_{r'p q'}^{(3)}
\right)
\right]_\Lambda^k, \label{1loopCSa} \\
\partial_t \Gamma_{k,\ q'q}^{(2)} &=& 
-(G\ k^2\, G)_{pp'}
\left(    S_{p'pqq'}^{(4)}
      - 2 S_{p'rq}^{(3)}\ G_{rr'}\ S_{r'pq'}^{(3)}
\right)\,. 
\label{1loopCSb}
\end{eqnarray}
Combining \eq{CS2-loop}, \eq{1loopCSa} and \eq{1loopCSb} leads us to  
\begin{eqnarray}\nonumber
\lefteqn{
\Delta\hat\Gamma_2 
=  
\int_\Lambda^k \s0{dk'}{k'}
\left\{ 
-\left[
G_{pp'}
\left(  S_{p'pqq'}^{(4)}
      - S_{p'rq}^{(3)}\ G_{rr'} \ S_{r'pq'}^{(3)}
\right)
\right]_\Lambda^k
(G\ k'^2 G\ k'^2\, G)_{q'q}
\right.}
\hspace{2.7cm}\\ \di \tab \tab 
\left.
 - \s012 (G\ k'^2\, G)_{pp'}
\left( 
   S_{p'pqq'}^{(4)}
 - 2 S_{p'rq}^{(3)}\ G_{rr'} \ S_{r'pq'}^{(3)}
\right) 
(G\ k'^2\, G)_{q'q}
\right\}
\label{CSfinal}\end{eqnarray}
We rewrite the integrand in \eq{CSfinal} in terms of total derivatives
with respect to the scale parameter $t$. Again a graphical
representation for the integrand is helpful, cf.~Fig.~15, where the
definitions of Fig.~2 and Fig.~4 have been used with $\partial_t
R=2k^2$ and $R=k^2$.

\begin{figure}
\begin{center}
\unitlength0.001\hsize
\begin{picture}(600,200)
\psfig{file=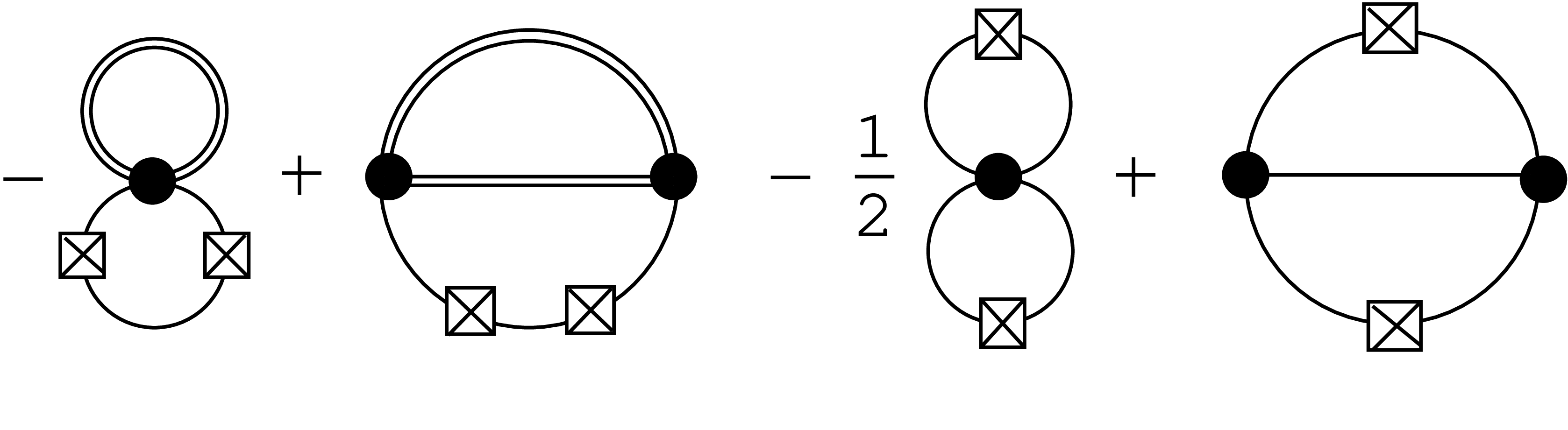,width=.6\hsize}
\end{picture}
\begin{minipage}{.8\hsize}{
{
  \small {\bf Figure 15:} The integrand in curly brackets of
  \eq{CSfinal}.} See Fig.~8 for comparison with the standard proper
time flow for $m=2$.}
\end{minipage}
\end{center}
\end{figure}
\noindent
Using Fig.~5, we rewrite Fig.~15 in terms of total derivatives.
Inserting the simple graphical derivative in \eq{CSfinal} we end up
with
\begin{eqnarray}\nonumber 
\Delta\hat\Gamma_2 
\tab = \tab 
\int_\Lambda^k \s0{dk'}{k'}
\left\{ 
\s014 \partial_{t'}\left( 
G_{pp'}\ 
(S_{p'pqq'}^{(4)} - S_{p'rq}^{(3)}\ G_{rr'}\ S_{r'pq'}^{(3)})\ 
(G\ k'^2\, G)_{q'q}
\right)
-{\rm subtractions}\right\}
\\ \di \nonumber 
\tab - \tab 
\int_\Lambda^k 
\s0{dk'}{k'}
\left\{
\s012
G_{pp'}(S_{p'pqq'}^{(4)}-\ S_{p'rq}^{(3)}\ G_{rr'}\ S_{r'pq'}^{(3)})\ 
(G\ k'^2\, G)_{q'q} -{\rm subtractions}\right\}
\\ \di
\tab =\tab 
\left[ \s018      G_{pp'} \ S_{p'pqq'}^{(4)}\ G_{q'q}
      -\s0{1}{12} G_{pp'} \ S_{p'rq}^{(3)} \ G_{rr'}\ S_{r'pq'}^{(3)} \
G_{q'q}\right]_{\rm ren.}
\label{CStotal} 
\end{eqnarray}
This is the correct two loop result as displayed in \eq{profield}. In
order to arrive at \eq{CStotal}, we made use of the fact that the
total derivative term in the first line of \eq{CStotal} vanishes at
$k=0$. The second line can be written as a total $t$-derivative by
noticing that in the present case $G\ k^2\ G = -\s012 \partial_t G$.
It reduces the second line of \eq{CStotal} to the first line of
\eq{profield}. This proof of perturbative completeness can be extended
to arbitrary high orders within the loop expansion.  \step

This offers an alternative way to arrive at the result \eq{deviation}.
We study the difference of the integrated flow \eq{CS2-loop} to the
integrated proper-time flow in \eq{PTRG2loop}. The difference between
the two flows is given by
\begin{eqnarray}\nonumber 
\Delta\Gamma_2-\Delta\hat\Gamma_2
&=&
- \int_\Lambda^k \s0{dk'}{k'}
\left\{
 (G\ k'^2\ G)_{pp'}
 \left(S_{p'pqq'}^{(4)} -2 S_{p'rq}^{(3)}\ G_{rr'}\ S_{r'pq'}^{(3)}\right)
 (G\ k'^2\ G\ k'^2\ G)_{q'q}
\right\}
\\ &&+
\int_\Lambda^k \s0{dk'}{k'}
\left\{
(G\ k'^2\ G)_{pp'}
\left(\s012 S_{p'pqq'}^{(4)}- S_{p'rq}^{(3)}\ G_{rr'}\ S_{r'pq'}^{(3)}\right) 
(G\ k'^2\ G)_{q'q}
\right\}
\label{difference}
\end{eqnarray} 
modulo subtractions. \Eq{difference} can also be deduced from the
recursive relation between $\Delta\Gamma_m$ and
$\Delta\hat\Gamma_{m-1}$ as displayed in App.~\ref{secrecursive},
eqs.~\eq{recursive} and \eq{recursive2}.  If the proper-time flow was
complete the difference would vanish since the CS flow is complete.
After some straightforward algebra this leaves us with the following
consistency condition:
\begin{eqnarray}\label{nonstandard}
0\stackrel{!}{\equiv} \Delta\Gamma_2-\Delta\hat\Gamma_2=
-\s012 \int_0^\infty \s0{dk}{k} 
(G\ k^2\ G)_{pp'}\ S_{p'rq}^{(3)}\
(G\ k^2\ G)_{rr'}\ S_{r'pq'}^{(3)}\ 
(G\ k^2\ G)_{q'q}\,,
\end{eqnarray} 
which is not satisfied. Using \eq{CStotal} and \eq{nonstandard} leads
us to the representation \eq{deviation} of the proper-time flow.

\section{Recursive relations}\label{secrecursive}

In this appendix, we derive two loop recursive relations for 
proper-time flows for values $m$ and $m'$ that differ by an integer. 
These relations make the scheme
dependent deviation from perturbation theory explicit. The result is
used in Sect.~\ref{SectionPTRG}.\step

The equation for the two loop contribution to a flow with parameter
$m$ is
\begin{eqnarray}\label{2-loopm}
\Delta\Gamma_{2,m} = -m \int_\infty^0 \s0{dk}{k}
\Tr (G\ k^2)^{m}\,G\, \Delta\Gamma_{1,m}^{(2)} .
\end{eqnarray}
 with $G=(S^{(2)}+k^2)^{-1}$. 
\Eq{2-loopm} can be rewritten in terms of $\Delta\Gamma_{2,m-1}$ and loop
terms. In the following it is understood that integrals between $k=0$ and
$k=\infty$ of total derivatives proportional to $k^2$ vanish up to
renormalisation. Now we use that
\begin{eqnarray}\label{id1}
 - m (G\ k^2)^{m}\, G=
\s012 \partial_t \left[(G\ k^2)^{m-1}\, G\right] -
(m-1)(G\ k^2)^{m-1}\,G.
\end{eqnarray}
Using also a partial integration we get from \eq{2-loopm} and \eq{id1}
 \begin{eqnarray}\label{2-loopm1}
\Delta\Gamma_{2,m} = -\s012\int_\infty^0 \s0{dk}{k}
\Tr (G\ k^2)^{m-1}\, G\,\partial_t\Delta\Gamma_{1,m}^{(2)}
-(m-1)\int_\infty^0 \s0{dk}{k}
\Tr (G\ k^2)^{m-1}\, G\,\Delta\Gamma_{1,m}^{(2)}.
\end{eqnarray}
If we could substitute $\Delta\Gamma_{1,m}^{(2)}$ by 
$\Delta\Gamma_{1,m-1}^{(2)}$ in the second term on the right-hand side of 
\eq{2-loopm1}, this term would just be $\Delta\Gamma_{2,m-1}$, 
as can be seen from \eq{2-loopm}. To that end notice that 
\begin{eqnarray}
\label{id3}
(G\ k^2)^{m}-(G\ k^2)^{m-1}=-(G\ k^2)^{m-1} G S^{(2)}
= \s0{1}{2(m-1)}\partial_t (G\ k^2)^{m-1}.
\end{eqnarray}
With \eq{id3}, it is possible to express the one loop contribution
$\Delta \Gamma_{1,m}$ in terms of $\Delta \Gamma_{1,m-1}$ and a one loop
term:
\begin{eqnarray}\label{id2}
\Delta \Gamma_{1,m}=\Delta \Gamma_{1,m-1}-\s0{1}{2(m-1)}
\left[\Tr (G\, {k'}^2 )^{m-1}\right]_\Lambda^k\,.
\end{eqnarray}
Alternatively, \eq{id2} can be read off from \eq{1,m}, or more
easily for integer $m$ from \eq{del1,m}.  Using \eq{id3} in the
second term on the right-hand side of \eq{2-loopm1}, this term takes
the form
\begin{eqnarray}
\Delta\Gamma_{2,m-1}
+\s012
\int_\infty^0 \s0{dk}{k}\ \Tr\ (G\ k^2)^{m-1}\,G\,
{\delta^2\over(\delta\phi)^2}\ \Tr\, (G \,{k}^2)^{m-1}\,.
\label{identify}\end{eqnarray}
Next, we consider the first contribution on the right-hand side of
\eq{2-loopm1}, where we use 
\begin{equation}\label{id4}
\partial_t\, \Delta\Gamma_{1,m}^{(2)}
={\delta^2\over(\delta\phi)^2} \ \Tr\ (G\,k^2)^m\,.
\end{equation}
Combining the first term in \eq{2-loopm1}, using \eq{id4}, with the
second term in \eq{identify}, and making use of the first equation in
\eq{id3}, we arrive at the recursive relation
\begin{eqnarray}
\Delta\Gamma_{2,m}-\Delta\Gamma_{2,m-1}=
\012 \int_\infty^0
\0{dk}{k}\,\Tr \left[({G\,{k}^2})^{m-1}\, G\,
\,{\delta^2\over(\delta\phi)^2}\Tr\, (G \,{k}^2)^{m-1} G\,S^{(2)}
\right]\,,
\label{recursive} \end{eqnarray}
apart from irrelevant terms from the different renormalisation
procedures for the two flows. \Eq{recursive} cannot be written as the
integral of a total derivative. We can, however, perform a partial
integration using
$\partial_t(\Delta\Gamma_{1,m}^{(2)}-\Delta\Gamma_{1,m-1}^{(2)})=-\Tr\,
(G \,{k}^2)^{m-1} G\,S^{(2)}$. Employing also \eq{id2}, we end up with
\begin{eqnarray}
\Delta\Gamma_{2,m}-\Delta\Gamma_{2,m-1}=
\012\int_\infty^0
\0{dk}{k}\,\Tr \left[({G\,k^2})^{m}\,
\,(\s0{m}{m-1}\, G-k^{-2})
\,{\delta^2\over(\delta\phi)^2}\Tr\, (G \,{k}^2)^{m-1}
\right]\,,
\label{recursive2} \end{eqnarray}
which has been given previously in \cite{Litim:2001ky}. The different
forms could prove useful when discussing the terms dropped in a
specific proper-time flow. Finally, \eq{recursive} can be used to
write down a general relation between flows with $m,m'$ that differ by
an integer $n$. We have
\begin{eqnarray}
\Delta\Gamma_{2,m}=\Delta \Gamma_{2,m-n}+
\012\sum_{l=m-n}^{m-1}\int_\infty^0
\0{dk}{k}\, \Tr \ \left[({G\,{k}^2})^{l}\, G\,
\,{\delta^2\over(\delta\phi)^2}\Tr\, (G \,{k}^2)^{l} G\,S^{(2)}
\right]\,.
\label{m-n} \end{eqnarray}
The difference \eq{m-n} depends on arbitrarily high powers of the
fields and does not integrate to zero. Similar relations also exist
for non-zero $k$, but then we also have contributions that integrate
to zero as they are total derivatives of terms proportional to $k^2$.


\end{document}